\documentclass[aps,prc,twocolumn,showpacs,groupedaddress]{revtex4}

\usepackage{graphics}
\usepackage{amsmath}
\usepackage{ulem}
\usepackage{comment}

\def\nuc#1#2{\relax\ifmmode{}^{#1}{\protect\text{#2}}\else${}^{#1}$#2\fi}
\newcommand{\etal}{\textit{et al.~}}

\newcommand{\be}{\begin{eqnarray}}
\newcommand{\ee}{\end{eqnarray}}

\begin{document}


\title{Particle motion in a deformed potential using a
  transformed oscillator basis} 



\author{J.~A. Lay}
\email{lay@us.es}
\author{A.~M.\ Moro}
\email{moro@us.es}
\author{J.~M.\ Arias}
\email{ariasc@us.es}
\affiliation{Departamento de FAMN, Facultad de F\'{\i}sica, Universidad de Sevilla, Apdo.~1065, E-41080 Sevilla, Spain}
\author{J.\ G\'omez-Camacho}%
\email{gomez@us.es}
\affiliation{Departamento de FAMN, Facultad de F\'{\i}sica, Universidad de Sevilla, Apdo.~1065, E-41080 Sevilla, Spain\\}
\affiliation{Centro Nacional de Aceleradores, Avda.\ Thomas A. Edison,
  E-41092, Sevilla, Spain} 

\vspace{1cm}

\date{\today}


\begin{abstract} 
The quantum description of a particle moving in a deformed potential
is investigated.
A pseudostate (PS) basis is used to represent the states of the
composite system. This PS basis is obtained by diagonalizing the
system Hamiltonian in a family of square integrable functions. In this
work the Transformed Harmonic Oscillator (THO) functions, obtained from the
solutions of the Harmonic Oscillator using a Local Scale
Transformation (LST), are used. The proposed method is applied to the
$^{11}$Be nucleus, 
treated in a two-body model ($^{10}{\rm Be}+n$). Both structure and
reaction observables have been studied.

Wavefunctions and energies obtained for the bound states and some
low-lying resonances are compared 
with those obtained by direct integration of the Schr\"odinger
equation. The dipole and quadrupole electric transition probabilities
for the low-energy continuum have been calculated in the THO basis,
and compared with the {\it exact} distributions obtained with the
scattering states.  
Finally, the method is applied to describe the  $^{11}$Be states
in the Coulomb breakup of  $^{11}$Be+$^{208}$Pb at 69~MeV/nucleon. The
energy and angular distributions of the exclusive breakup have been
calculated using the Equivalent Photon Method, including both E1
and E2 contributions. The calculated distributions are found to be in
good agreement with the available experimental data from RIKEN [Phys. Rev. C70, 054606]. At 
the very forward angles, the cross section is
completely dominated by the dipole couplings. 

\end{abstract}

\maketitle

\section{\label{intro} Introduction}
It is well known that the quantum collision of a weakly bound system
by a target is influenced by the coupling to the unbound states of  
the projectile. For nuclear collisions, this effect was first noticed
in deuteron-induced reactions, and later observed in the scattering of
other loosely bound nuclei, such as halo nuclei.  Several reaction
frameworks have been envisaged to account for this effect. Among them,
one of the most successful has been the Continuum-Discretized
Coupled-Channels (CDCC) method \cite{Raw74,Aus87}, originally
developed to account for the breakup channels in deuteron  
scattering and later extended to other weakly-bound nuclei, such as
$^{6,7}$Li, $^{11}$Be, or $^{8}$B, among others. In all these cases,  
the projectile is described in a two-body model ($^6$Li=$^4$He+$d$,
$^7$Li=$^4$He+$^3$H, $^{11}$Be=$^{10}$Be+$n$, etc) and the method  
considers explicitly the possible  dissociation of the projectile into
its two fragments. In its standard formulation, the excitation of each
fragment is nevertheless ignored.   This is a good
approximation for deuteron scattering, for which both constituents can
be considered inert at the  
energies of interest in nuclear studies, but it is more questionable
for more complex systems. Moreover, the bound and unbound states of the
two-body system are considered to be well described by pure
single-particle configurations. This approximation ignores possible
admixtures of different core states in the wave functions of the
projectile. These admixtures are known to be important, particularly
in the case of well-deformed cores, as for example in the  $^{11}$Be
halo nucleus.

A recent attempt to accommodate these effects within the CDCC method
was done in Ref.~\cite{Summers06}, and applied to the scattering of
one-neutron halo nuclei. In that work, the states of the core+valence
system were described within the particle-rotor model                       
\cite{BM}. The unbound states of the compound  system were described
by continuum bins which, following the standard procedure,  
were constructed by superposition of scattering  states. These
scattering states are obtained by direct integration of the
Schr\"odinger equation, with the  
appropriate boundary conditions.

Alternatively, the bound and unbound states of the system can be
obtained by diagonalizing the Hamiltonian in a suitable basis of
square-integrable functions. The eigenfunctions of the system are
expressed as an expansion in the basis functions. In practical
calculations, the basis needs to be truncated, leading to a finite
expansion   
of the eigenfunctions. Therefore, these states and their corresponding
eigenvalues can be regarded as a finite approximation to the exact
states of  
the system and are referred hereafter as pseudo-states (PS). This
procedure has been applied, for example, to describe the states of
two-body nuclei interacting via a  
central potential \cite{Mat03,Per02,Mor09}
and, more recently, also for three-body nuclei
\cite{Mat04a,Mat04b,Mat06,manoli08}. A variety of bases have been used
in these applications, such as harmonic oscillator (HO), Gaussian,
Laguerre functions, etc. The procedure can be also extended to
deformed systems. A natural choice for the PS would be the deformed HO
potential   
\cite{Vau73,Gam90}. However, this basis is not suitable to describe
the bound states of weakly-bound nuclei due to its Gaussian
asymptotic behavior. Several alternatives have been proposed in the
literature, for example, the eigenstates of a  truncated Woods-Saxon
potential \cite{Zho03} or the Sturmian basis \cite{Ban80,Nun96}.

In this work, we propose the use of a Transformed Harmonic Oscillator
(THO) basis to describe the states of a two-body system mutually interacting
with a deformed potential.  This basis has been previously applied to
the case of spherical systems \cite{Mor09} so we present here its
extension to deformed systems. The THO basis is obtained by  
applying a Local Scale Transformation (LST) to the Harmonic Oscillator
(HO) basis. The LST, adopted from a previous  
work of Karataglidis \etal \cite{Amos}, is such that it transforms the
Gaussian asymptotic behavior into an exponential form,  thus ensuring
the correct asymptotic behavior for the bound wave functions. The
accuracy of this THO basis was tested for several  
reactions induced by deuteron and halo nuclei, showing an excellent
agreement with the standard binning method, and an improved
convergence rate in the case of narrow resonances \cite{Mor09,Lay10}.  

For a deformed potential, the calculation of bound and unbound states
becomes a multi-channel problem, since, in general, for each physical
state there will be contributions from several orbital angular
momenta and core states. For bound states, the calculation of the
energies and eigenfunctions  is analogous to the
single-channel case, because these quantities are directly obtained
from the diagonalization in the chosen PS basis. For unbound states,
the eigenfunctions (and their corresponding eigenvalues) obtained from
the Hamiltonian diagonalization can be regarded as a finite and
discrete representation of the {\it exact} states. In general,
resonances (quasi-stationary states) correspond to combinations of
these positive-energy eigenstates and hence its identification is not
straightforward.  

For a particle moving in a central potential (with possibly a
spin-orbit component) this is a relatively  
straightforward problem and indeed a variety of methods have been
proposed to compute resonance energies and widths. For example, they
can be obtained from the poles  
of the $S$-matrix in the complex energy plane. A simpler method is to
define the resonance as the energy at which the phase-shift
crosses $\pi/2$. The width is then obtained from the inverse of the
derivative of the phase-shift, evaluated at the energy of the
resonance. These methods rely on the knowledge of the scattering states at
large distances (from which the $S$-matrix and hence the phase-shifts
can be extracted)  and then they cannot be directly applied to PS
methods, given the wrong asymptotic  
behaviour of the PS functions. In this case, the identification of resonances can be
done using the so-called  
{\it stabilization method} \cite{Haz70,Tay76}. This is a procedure
envisaged to identify and construct the most localized continuum wave
functions when the positive  
energy states are expanded in a discrete basis, depending on one or
more parameters.  In practice, this can be achieved by  diagonalizing
the Hamiltonian as a function of these parameters (for example, the
basis size) and then scanning the resultant eigenvalues for the
continual appearance of a stabilized value which, unlike the others,  
is insensitive to the size of the basis. In some previous works, we
have successfully applied this technique to obtain the resonances of
two-body systems with central potentials using the THO basis
\cite{Mor09,Lay10}. In this work, we explore the validity of this
method for the multi-channel situation that arises in the deformed
case.  Our aim with this work is to assess the capability of the THO
basis for calculations including core deformation in the simpler
two-body systems, as $^{11}$Be. This step is necessary and unavoidable
for providing a solid basement to proceed with the generalization of
the formalism to more challenging situations, such as the case of three-body composite
systems including core deformation, or to the scattering of a two-body system by a third body, 
including core deformation in one of the clusters of the composite system. 

The work is structured as follows. In Section \ref{sec:thoamos}
the THO method based on the parametric LST is reviewed and 
the structure model used in subsequent
calculations is discussed. In Section \ref{sec:bel}, general expressions
for the electric transition operators for the particular case of a
two-body system with a deformed core are provided. In Sec.~\ref{sec:structure} the
model is applied to describe the structure of the $^{11}$Be
nucleus. The basis so obtained is then used to describe the Coulomb
breakup of $^{11}$Be on $^{208}$Pb at 69~MeV/nucleon,
comparing our results with the available data. Finally, in Section
\ref{sec:summary} the main results of this work are summarized.

\section{\label{sec:thoamos}  Eigenstates of a deformed potential in a
  PS basis: the THO basis }  
In this section, we briefly review the features of the PS basis used
in this work. This basis is an extension of the THO basis used in our
previous  
works to describe the states of a composite system consisting of two
interacting inert fragments, such as a valence particle
(proton/neutron) and a spherical and stable core. The goal of this
extension is to allow core-excited admixtures in the description of
the  
states of the composite system and hence the possibility of dynamic
core excitation mechanisms in reactions involving these nuclei. For
completeness, we  review first the situation in which the core degrees
of freedom are neglected. In this case,  the core+valence Hamiltonian
is simply given by: 
\be
H= T_r + V_{vc}(\vec{r}) 
\label{hsp}
\ee
where $\vec{r}$ is the relative coordinate between the valence and the
core, $T_r$ the core-valence kinetic energy operator and  
$V_{vc}(\vec{r})$ is the interaction between the valence particle and the core. 
The eigenstates of this Hamiltonian can be characterized by the
energy eigenvalues ($\varepsilon$) and the set of quantum numbers
$\{\ell, s, j \}$, which correspond to the orbital angular momentum
($\ell$), the valence spin ($s$) and their sum ($\vec{j}=\vec{\ell} +
\vec{s}$). For a central potential with, possibly, a spin-orbit term,
these states can be written as: 
\begin{equation}
\phi_{\varepsilon,\ell,j}(\vec{r}) = R_{\varepsilon,\ell,j}(r)  {\cal
  Y}_{\ell s j m}(\hat{r})   
\label{wfgen}
\end{equation}
where  ${\cal Y}_{\ell s j m}(\hat{r}) = \left[ Y_{\ell}(\hat{r})
  \otimes \chi_s \right]_{jm}$, with  $\chi_s$  a spin function. The
radial functions $R_{\varepsilon,\ell j}(r)$ can be obtained by solving the
Schr\"odinger equation subject to the appropriate boundary condition
for bound ($\varepsilon < 0$) or unbound ($\varepsilon > 0$)
states. Alternatively, these functions  can be obtained by
diagonalizing the Hamiltonian   (\ref{hsp})  in a discrete
basis. Since any complete basis  will be infinite, this procedure is
not feasible in practice unless the basis is  truncated. By doing so,
one obtains a finite (an approximate) expansion of the functions
$R(r)$ in the selected basis. If the basis  
functions are denoted by $\varphi_{n,\ell,j}(\vec{r})= \chi_{n,\ell}(r) {\cal
  Y}_{\ell s j m}(\hat{r})$, we will have: 
\begin{equation}
R_\beta (r)= \sum_{n=1}^{N}  c_{\beta,n} \chi_{n,\ell}(r) 
\end{equation}
where $\beta \equiv \{\varepsilon,\ell,s,j\}$ and  $N$ is the number
of states retained in the basis.  

As already mentioned, there are many possible choices for
the  basis functions $\{\varphi_n\}$ (Gaussians, harmonic oscillator,
Laguerre, etc). In this work we use the transformed
harmonic oscillator (THO) basis, obtained from the harmonic oscillator
basis with an appropriate LST \cite{SP88,PS91}.
If the LST function is  
denoted by $s(r)$, the THO states are obtained as 
\begin{equation}
\label{eq:tho}
R ^{THO}_{n, \ell}(r)= \sqrt{\frac{ds}{dr}} R^{HO} _{n, \ell}[s(r)],
\end{equation}
where $R ^{HO}_{n, \ell}(s)$ is the radial part of the HO functions. 
With the criterion given above, the LST is indeed not unique. In
Ref.~\cite{Per01} the LST was defined in such  
a way that the first HO state is exactly transformed into the exact
ground state wave function, assuming that this is known. Therefore, by  
construction, this wave function is exactly recovered for any
arbitrary size of the basis. In a more recent  
work \cite{Mor09} we adopted the parametric form of Karataglidis \etal
\cite{Amos} 
\begin{equation}
\label{lstamos}
s(r)  = \frac{1}{\sqrt{2} b} \left[  \frac{1}{   \left(  \frac{1}{r}
    \right)^m  +  \left( 
\frac{1}{\gamma\sqrt{r}} \right)^m } \right]^{\frac{1}{m}}\ ,
\end{equation}
that depends on the parameters $m$, $\gamma$ and the oscillator length
$b$. Note that,  
asymptotically, the function  $s(r)$ behaves as 
$s(r)\sim \frac{\gamma}{b} \sqrt{\frac{r}{2}}$
and hence the functions obtained by applying this LST to the HO basis
behave at  
large distances as $\exp(-\gamma^2 r / 2 b^2)$. Therefore, the ratio
$\gamma/b$ can be related to an effective linear momentum, 
$k_\mathrm{eff}=\gamma^2 /2 b^2$, which  
governs the asymptotic behaviour of the THO functions. As the ratio
$\gamma/b$ increases, the radial extension of the basis decreases and,  
consequently, the eigenvalues obtained upon diagonalization of the
Hamiltonian in the THO basis tend to concentrate at higher excitation
energies. Therefore, $\gamma/b$ determines the density of eigenstates
as a function of the excitation energy. In all the calculations
presented in this work, the power $m$ has been taken as $m=4$. This
choice is discussed in  Ref.~\cite{Amos} where the authors found that
the results are weakly dependent on $m$.

Note that, by construction, the family of functions  
\( R ^{THO}_{n, \ell}(r) \) are orthogonal
and constitute a complete set with the following normalization: 
\begin{equation}
 \int_{0}^{\infty} r^2 | R ^{THO}_{n, \ell}(r)|^{2} dr=1  \, .
\end{equation} 
Moreover, they decay exponentially
at large distances, thus ensuring the correct asymptotic behaviour
for the bound wave functions. In practical calculations a finite set
of functions (\ref{eq:tho})  
is retained, and the internal Hamiltonian of the projectile is
diagonalized in this truncated basis with $N$ states,  
giving rise to a set of  eigenvalues and their associated
eigenfunctions, denoted respectively by $\left\{\varepsilon_n \right\}$ and 
$\{\varphi^{(N)}_{n, \ell}(r)\}$ ($n=1,\ldots,N$). As the basis size
is increased,  the eigenstates with negative energy
will tend to the exact bound states of the system, while those
with positive eigenvalues can be regarded as a finite representation
of the unbound states. 

The formalism can be extended to the situation in which  the core
degrees of freedom are taken into account explicitly. In this
case, the Hamiltonian (\ref{hsp}) is  generalized to 
\be
H= T_r + V_{vc}(\vec{r},\vec{\xi}) + h_{\rm core}(\vec{\xi}) 
\label{hpc}
\ee
where $h_{\rm core}(\vec{\xi}) $ is 
the intrinsic Hamiltonian of the core, whose eigenstates will be
denoted by $\{\phi_{I M_I} \}$. 
Additional quantum numbers, required to fully specify the core states,
are not included for notation simplicity. Note that the
valence-target interaction, $ V_{vc}(\vec{r},\vec{\xi})$, contains now
a dependence on the core degrees of freedom (denoted generically by
$\vec{\xi}$).  

The eigenstates of the Hamiltonian cannot any longer be written in the
form of Eq.~(\ref{wfgen}). Instead, these states  
will be a superposition of several valence configurations and core
states, i.e. 
\be
\Psi_{\varepsilon; J M }(\vec{r},\vec{\xi}) 
 =  \sum_{\alpha} R_{\varepsilon,\alpha}(r) 
\left[  {\cal Y}_{\ell s j m}(\hat{r}) \otimes \phi_{I}(\vec{\xi}) \right]_{JM} .
\label{wfx}
\ee 
Upon replacement of the expansion (\ref{wfx}) into the Schr\"odinger
equation, one gets a coupled set of differential equations for the
radial functions $R_{\varepsilon,\alpha}(r)$. For bound states, these radial
functions decay exponentially for $r\rightarrow \infty$ giving rise to
square-integrable functions.  For continuum states, the functions
$R_{\varepsilon,\alpha}(r)$ are also  
obtained by solving a set of coupled radial equations, but subject to
the boundary condition that incident  
waves occur only in the entrance channel characterized by a given set
of quantum numbers $\alpha=\{ \ell,s,j,I \}$.   
Therefore, for each continuum energy, there are as many scattering
solutions as possible  
values of  $\alpha$, compatible with the total angular momentum $J$.

Alternatively, the functions $R_{\varepsilon,\alpha}(r)$ can be
obtained using an expansion in a PS basis, such as the THO basis
described above. In this case, the basis must include also the new
core degree of freedom 
\begin{equation}
\Phi^\alpha_{n,J M }(\vec{r},\vec{\xi}) 
 =  R^{THO}_{n,\alpha}(r) \left[  {\cal Y}_{\ell s j m}(\hat{r}) \otimes
   \phi_{I}(\vec{\xi}) \right]_{JM} . 
\label{basis2}
\end{equation}

In this basis, the states of the system will be expressed as
\begin{equation}
\Psi^{(N)}_{i,J M }(\vec{r},\vec{\xi}) 
 =  \sum_{n=1}^{N} \sum_{\alpha} c^i_{n,\alpha,J} \Phi^\alpha_{n,J M
 }(\vec{r},\vec{\xi}) ,
\end{equation}
where $i$ is an index that labels the order of the eigenstate. 


These eigenstates are spread in the energy spectrum with a density strongly related to  the basis parameters, mainly  $N$ and $\gamma / b$, and to the continuum structure for the selected Hamiltonian, i.e.\ presence of resonances or different breakup thresholds. Moreover, this density reflects the momentum distribution of the eigenstates which becomes important to obtain 
continuous energy or momentum distributions of different observables from their discrete representation in the PS basis \cite{Mat03,Tos01,Mor09,Lay10}. Generalizing the expression in \cite{Lay10}, the density of states is here defined as:
\begin{equation}
\rho (k)=\sum_{i=1}^{N}\sum_{\alpha}^{n_{\alpha}}\langle  k_{\alpha} J_f  | \Psi^{(N)}_{i,J M } \rangle,
\label{densk}
\end{equation}
where $| k_{\alpha} J_f  \rangle $ denotes the exact scattering wavefunction for an incoming wave in the $\alpha$ channel. Note that the difference between $k$ and $k_{\alpha}$ relays on the threshold energy for each channel.

With this definition the integral of the density with respect to the momentum is the number of THO functions selected (N) times the number of channels ($n_\alpha$):
\begin{equation}
\int_{0}^{\infty}\rho (k) \,dk=N  n_{\alpha},
\end{equation}
assuming that we have included N  THO functions for each channel $\alpha$. Note that this integrated density is independent of the LST parameters.

%
%
The afore-mentioned method can be applied to any Hamiltonian of the
form (\ref{hpc}). In the calculations presented in this work,  
the composite system is treated within  the particle-rotor model
\cite{BM}. Therefore,  we assume that the core
nucleus has a permanent deformation  which, for simplicity, is taken
to be axially symmetric. Thus, we can characterize the deformation by
a single parameter $\beta_2$.  In the body-fixed frame, the  
surface radius is then parameterized as $R(\hat{\xi})=R_0 [1 + \beta_2
  \, Y_{20}(\hat{\xi})]$, with $R_0$ an average radius. Starting from a central potential, $V^{(0)}_{vc}(r)$,
the full valence-core interaction is  obtained by deforming this  interaction as, 
\be
V_{vc}(\vec{r},\hat{\xi})=V_{vc}^{(0)}(r-\delta_2 Y_{20}(\hat{\xi})) ,
\label{Vvc}
\ee
with $\delta_2 = \beta_2 R_0 $ being the deformation
length. Transforming to the space-fixed frame of reference, and
expanding in  spherical harmonics, this {\it deformed} potential reads 
\be
V_{vc}(\vec{r},\vec{\xi}) = \sum_{{\cal L,M}}V_{vc}^{(\cal L)}(r)  
Y_{ {\cal L} {\cal M} }(\hat{r})
Y^{*}_{ {\cal L} {\cal M} }(\hat{\xi})~,
\label{vdef_vc}
\ee 
where the radial  form factors $V_{vc}^{(\cal L)}(r)$ are obtained by
projecting the deformed potential (\ref{Vvc}) onto the  
required multipoles. 


\section{\label{sec:bel} Electric transition probabilities in the PS basis} 
The accuracy of the PS basis to represent the continuum can be studied
by comparing the ground-state to continuum transition probability due
to a given operator. Here we consider the important case of the
electric dissociation of the initial nucleus into the fragments  
$c+v$.  This involves a matrix element between a bound state
(typically the ground state) and the continuum states.  

The electric transition probability between two bound states $| J_i
\rangle$ and $| J_f \rangle$ (assumed here to be unit normalized) 
 is given by the reduced matrix element (according to Brink and Satchler convention \cite{BS}) 
\begin{equation}
 \label{bediscgen}
{\cal B}(E\lambda; i \to f)=\frac{2 J_f+1}{2 J_i+1}\left | 
\langle J_f || \mathcal{M}(E\lambda) || J_i \rangle      \right |^2    ,
\end{equation}
where $\mathcal{M}$ is the multipole operator. In a core+valence
model, the electric transition operator can be written  
as a sum of three terms \cite{lay10a}: one for the excitation of the valence
particle outside the core, one for the excitation of the core as a
whole and one for mixed excitations involving simultaneous excitations
of core and valence particle, 
\begin{eqnarray}
\label{mel}
\mathcal{M}(E\lambda \mu) & = &   \sum_{k=1}^{\lambda-1}\sum_{m=-k}^{k}f_{\lambda}(k,m,\mu) 
\nonumber \\
&\times & \mathcal{M}_{sp}(E k m)  \mathcal{M}_{core}(E(\lambda-k) (\mu-m) )  \nonumber \\    
  &+ &  \mathcal{M}_{sp}(E\lambda \mu)  + \mathcal{M}_{core}(E\lambda\mu)  ,    
\end{eqnarray}
where $f_{\lambda}(k,m,\mu)$ is a well-defined function of its indices and the single particle contribution has the usual form,
\begin{equation}
\mathcal{M}_{sp}(E\lambda \mu)= Z_\mathrm{eff}^{(\lambda)} e r^\lambda
Y_{\lambda \mu}(\hat{r}), 
\end{equation}
with the effective charge:
\begin{equation}
Z_\mathrm{eff}^{(\lambda)}=Z_v
\left(\frac{m_c}{m_v+m_c}\right)^\lambda + Z_c
\left(-\frac{m_v}{m_v+m_c}\right)^\lambda . 
\end{equation}

In the case of a transition to a continuum of states,  
$|k J_f \rangle$, the  definition (\ref{bediscgen}) is  replaced by
(see for example \cite{Typ05}): 
\begin{equation}
\label{becontgen}
\frac{d{\cal B}(E\lambda)}{d \varepsilon}   =  \frac{2 J_f +1 }{2
  J_i+1} \frac{\mu_{vc} k }{(2 \pi)^3 \hbar^2}    
  \left | \langle k J_f || \mathcal{M}(E\lambda) || J_i \rangle \right |^2  , 
\end{equation}
with $k=\sqrt{2 \mu_{bc}\varepsilon}/\hbar$. Note that the extra 
factor appearing in Eq.~(\ref{becontgen}) with respect to
Eq.~(\ref{bediscgen}) is consistent with the convention $\langle k J
| k' J \rangle = \delta(k-k')$ and the asymptotic behaviour, 
\begin{eqnarray}
\label{wfasym}
u_{\alpha'}({k_{\alpha'}}, {r}) & \xrightarrow{r\rightarrow\infty} &   
 \frac{1}{2}i e^{2i\sigma_{l'}} \Big[ \delta_{\alpha' \alpha} H_{l}^{*}(k_{\alpha} r)
\nonumber   \\
& -&  \left( \frac{v_{\alpha}}{v_{\alpha'}}\right)^{\frac{1}{2}}
S^{(J)}_{\alpha',\alpha}H_{l'}(k_{\alpha'}r) \Big] ,  
\end{eqnarray}
where $u_\alpha(k_\alpha,r)= R_\alpha(k_\alpha,r) r$ (using an obvious
notation where the continuum $\varepsilon$ label has been replaced by
a dependence on the corresponding momentum $k$).

Using a finite basis, one may calculate only discrete values for the
transition probability.  According  
to Eq.~(\ref{bediscgen}), the $B(E\lambda)$ between the ground state (with angular momentum $J_i$) 
and the  $n$-th PS is given by
\begin{equation}
\label{beps}
{\cal B}^{(N)}(E\lambda; \mathrm{g.s.} \to n)=\frac{2 J_f+1}{2 J_i+1}
\left |
 \langle \Psi^{(N)}_{n,J_f} || \mathcal{M}(E\lambda) || \Psi_\mathrm{g.s.} \rangle  
\right |^2  . 
\end{equation}
In order to relate this 
discrete representation to the continuous distribution
(\ref{becontgen}) one may derive a continuous approximation to
(\ref{becontgen}) by introducing the identity in the truncated PS
basis, i.e. 
\begin{equation}
\label{closure}
I^{(N)}_{JM} = \sum_{n=1}^{N} | \Psi^{(N)}_{n,JM} \rangle  \langle  \Psi^{(N)}_{n,JM} |  .
\end{equation}
For $N \to \infty$ this expression tends to the \textit{exact}
identity operator for the Hilbert space spanned by the eigenfunctions
of the considered  
Hamiltonian. By inserting (\ref{closure}) into the exact expression
(\ref{becontgen}) we obtain the approximate continuous distribution, 
\begin{eqnarray}
\label{befold}
\frac{d{\cal B}(E\lambda)}{d\varepsilon} & \simeq  & \frac{2 J_f +1
}{2 J_i+1}  \frac{\mu_{vc} k }{(2 \pi)^3 \hbar^2} \nonumber \\  
& \times &  \left | \sum_{n=1}^{N}\langle  k J_f  | \Psi^{(N)}_{n,J_f} \rangle  
\langle  \Psi^{(N)}_{n,J_f} || \mathcal{M}(E\lambda) ||
\Psi_\mathrm{g.s.} \rangle   \right |^2 . \nonumber \\ 
\end{eqnarray}

This approach provides a {\it smoothing} procedure  to extract
continuous distributions, as a function of the asymptotic energy
$\varepsilon$ (or, equivalently,  
the linear momentum $k$), from the discrete distributions obtained
with the PS basis \cite{Mor06,Mor09}.  This is particularly convenient  
in situations in which the calculation with the scattering states
themselves is not possible, such as in the CDCC method.   

\section{\label{sec:structure} Test example: application to \nuc{11}{Be}}

\subsection{Energy spectrum and wave functions in the PS basis}

As an illustration of the formalism presented in the preceding
section, we consider the $^{11}$Be nucleus. This choice is motivated  
by the fact that this nucleus is one of the best known one-neutron
halo nuclei. Many of its properties can be understood in a simple  
two-body model, comprising a valence neutron orbiting a $^{10}$Be
core. For example, the ground state ($1/2^+$) and the only bound excited state ($1/2^-$)
are reasonably well described by $2s_{1/2}$ and $1p_{1/2}$ single-particle configurations, 
relative to the $^{10}{\rm Be}({\rm g.s.})$ core.
Excited states in the  continuum are also reasonably well 
described in terms of single-particle excitations of the halo neutron
outside the $^{10}{\rm Be}({\rm g.s.})$ core. This single-particle  
picture has been extensively used in the literature to explain also
reactions induced by this nucleus (see for instance
\cite{Cap04,How05,Lay10}). However, there are also numerous
experimental and theoretical evidences that these low-lying states of
$^{11}$Be contain significant  
admixtures of core-excited components
\cite{For99,Win01,Cap01,Cre11}. Consequently, an accurate description
of reactions involving  
this nucleus requires the inclusion of its states beyond the simple
single-particle picture.  


In the calculations presented in this work, we use the particle-rotor
model of Bohr and Mottelson with the $^{11}$Be Hamiltonian of Ref.~\cite{Nun96} (model Be12-b), which consists of a
Woods-Saxon central part, with a fixed geometry ($R=2.483$~fm,
$a=0.65$~fm) and a  parity-dependent strength. The potential contains
also a spin-orbit part, whose radial dependence is given by the
derivative of the same Woods-Saxon shape, and strength
$V_{so}=8.5$~MeV. For the $^{10}{\rm Be}$ core, this model assumes a
permanent quadrupole deformation   
$\beta_2$=0.67. Only the ground state ($0^+$) and the first excited
state  ($2^{+}$, $E_x= 3.368$ MeV) are included in the model
space. For the valence-core orbital angular momentum, we consider
the values $\ell \leq 3$.  

To generate  the THO basis we use the LST of Eq.~(\ref{lstamos}) with
$m=4$, $b=1.6$~fm and $\gamma=1.84$~fm$^{1/2}$. The value of  
$b$ was determined in order to minimize the ground state energy of
$^{11}$Be in a small THO basis. The factor $\gamma/b$ leads to a
$k_{eff}$ compatible with a maximum excitation energy of about 10~MeV,
which is enough for the calculations presented below.


\begin{figure}
{\par\centering \resizebox*{0.45\textwidth}{!}{\includegraphics{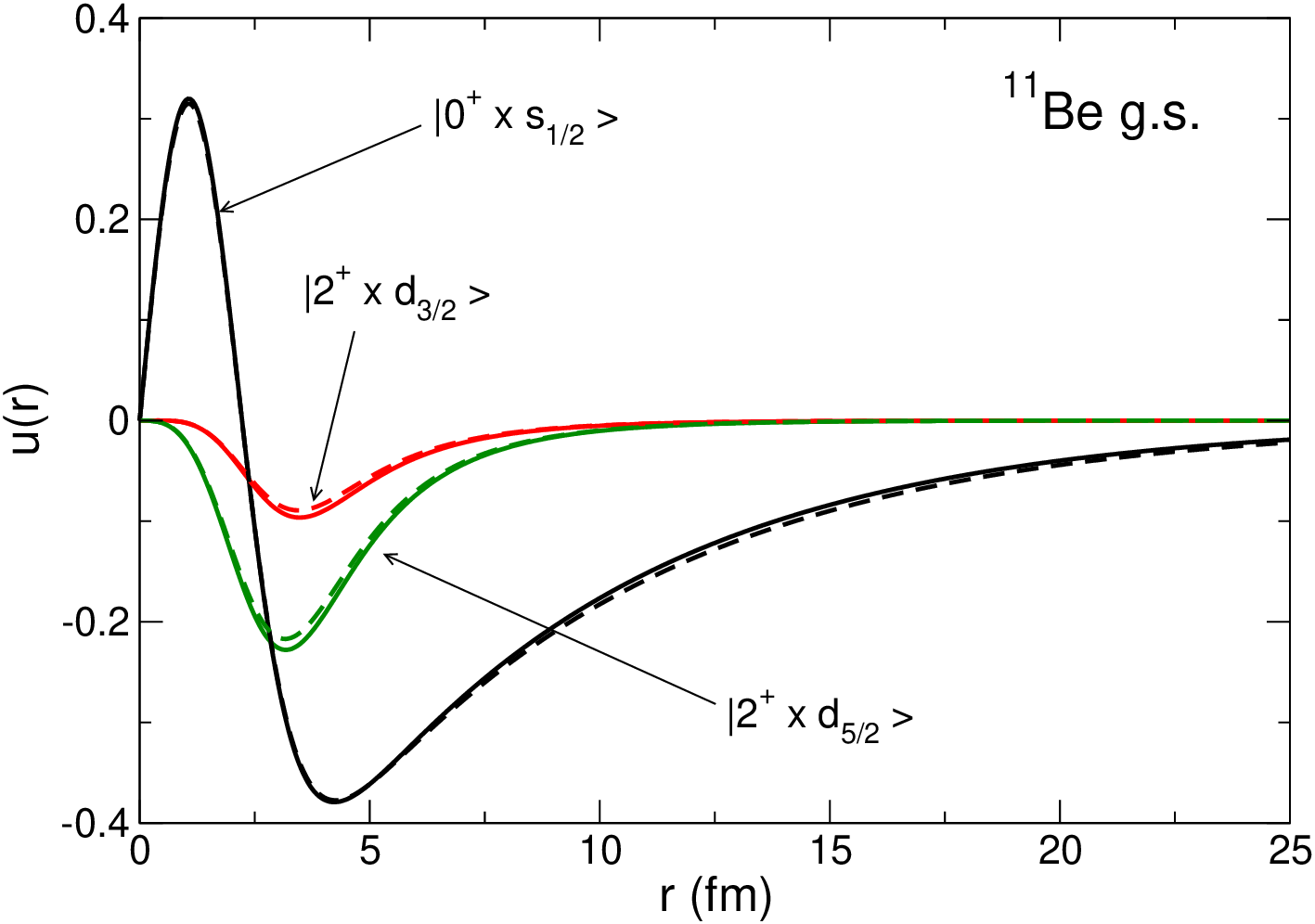}}\par}
 \caption{\label{wfgs} (Colour online) Radial parts of the ground
   state wave function for  the \nuc{11}{Be} nucleus obtained by
   direct integration of the Schr\"odinger equation (solid lines) and
   by diagonalization in a THO basis with N=15 states (dashed
   lines).} 
\end{figure}

Once these parameters have been fixed, the THO basis is generated for
different values of N, the number of oscillator functions, and  
the convergence of different observables is studied with respect to
this number. We should remark that the total number of basis functions
is this number times the number of channels $n_\alpha$. However, the latter 
 depends on the total angular momentum $J$ of the state under consideration, and will be the
same in any method based on the angular momentum expansion of the wave
functions. Therefore, we will refer to N as the basis size as we
understand it is the most honest way of comparing with other
methods. We find that the ground-state energy is already  
fully converged with a relatively small basis  (N $\approx 15$).

Within the model space used in our calculations ($I=0,2$,
$\ell \leq 3$), there are $n_\alpha=3$ channels contributing to the ground
state wave function, namely  $|^{10}{\rm Be}(0^+)
\otimes s_{1/2}\rangle$, $|^{10}{\rm Be}(2^+) \otimes d_{3/2}\rangle$
and  $|^{10}{\rm Be}(2^+) \otimes d_{5/2}\rangle$. 
In Fig.~\ref{wfgs}, we depict these radial parts of the ground-state 
wave function obtained from the diagonalization of the Hamiltonian in
a THO basis with N=15 oscillator functions (dashed lines). For
comparison, we include also the  
solutions obtained by direct integration of the  Schr\"odinger
equation (solid lines). Both calculations give basically identical
results. It can be seen, as expected, that the
$|^{10}{\rm Be}(0^+) \otimes s_{1/2}\rangle$ component is the dominant
one, accounting for about 80\% of the norm. This radial component
exhibits a node, due to the presence of a Pauli forbidden state
(arising from   the $1s_{1/2}$ orbital in the spherical basis).    

\begin{figure}
{\par\centering
  \resizebox*{0.45\textwidth}{!}{\includegraphics{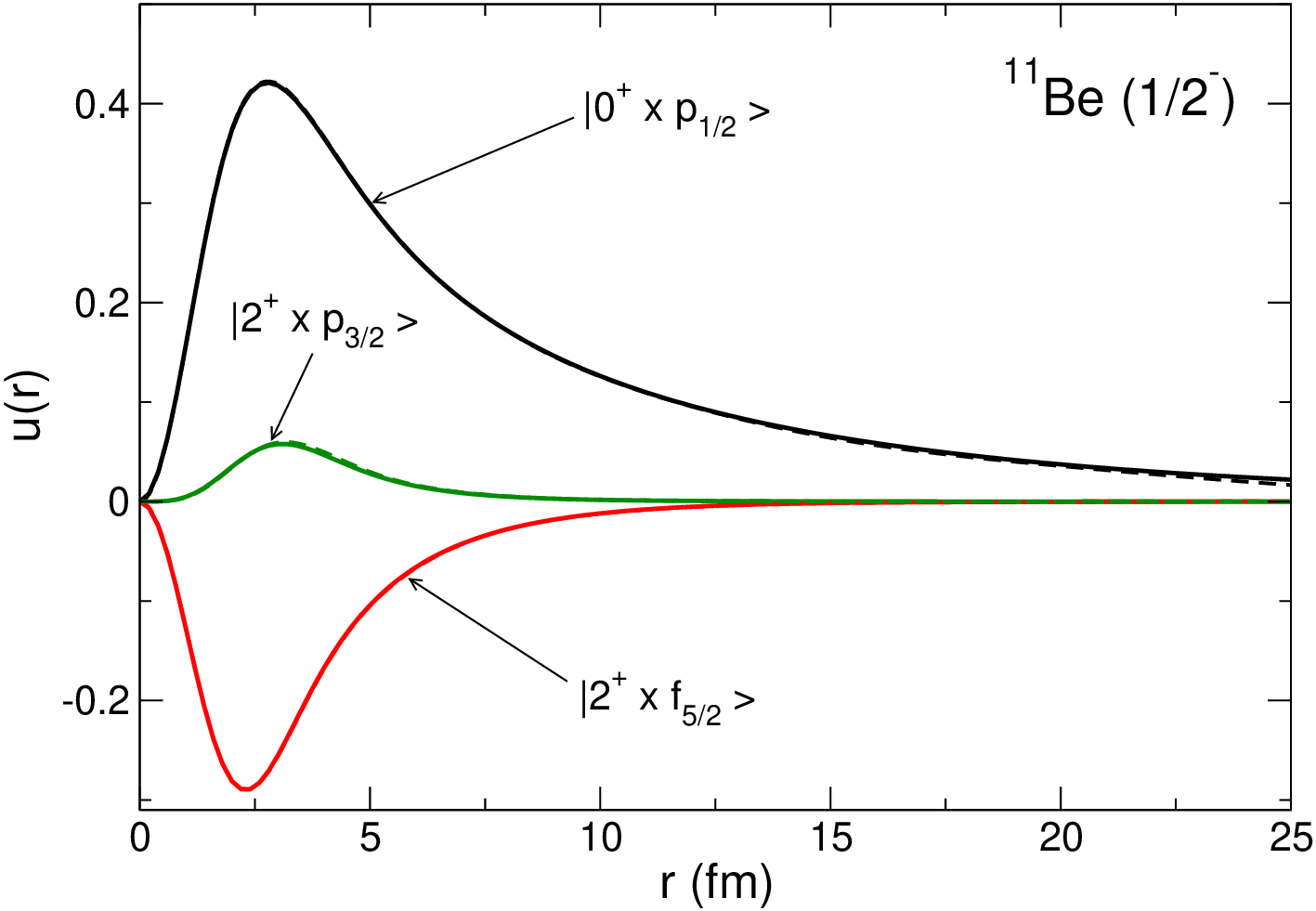}}\par} 
 \caption{\label{wf1ex} (Colour online) Radial parts of the 1/2$^-$
   excited bound state wave function for the \nuc{11}{Be} nucleus obtained by
   direct integration of the Schr\"odinger equation (solid lines) and
   by diagonalization in a THO basis with $N=15$ states (dashed
   lines).} 
\end{figure}

The assumed Hamiltonian reproduces also the position of the bound
excited state at $E_x=320$~keV ($1/2^-$). Indeed, this state appears
also in the diagonalization of the THO basis. The separation energy is
reproduced within a few percent with a basis of N=15 states and the
radial components are also found to be in perfect agreement with those
obtained by direct integration of the coupled differential
equations. This is shown in  Fig.~\ref{wf1ex}.  

We proceed to
discuss now the description of resonances in the PS basis. As
explained in the introduction, the identification of the resonances is
done using the stabilization method of Hazi and Taylor
\cite{Haz70,Tay76}, extended to the multi-channel case. The procedure
is the same as in the single-channel case, i.e., we diagonalize the
Hamiltonian over either a successively larger basis set or as a
function of a continuous parameter which defines the basis for a given
N value. Then, the evolution of the spectrum as a function of N or
the continuous parameter is studied. When a resonance is
present, there are some eigenvalues whose energies  
are stabilized for a range of values of N or the continuous
parameter. This property has been employed empirically in many works,
and a formal justification has also been provided by Lippmann and
O'Malley \cite{Lip70}. 

\begin{figure}
\begin{minipage}[b]{0.95\linewidth}
{\par\centering
  \resizebox*{0.95\textwidth}{!}{\includegraphics{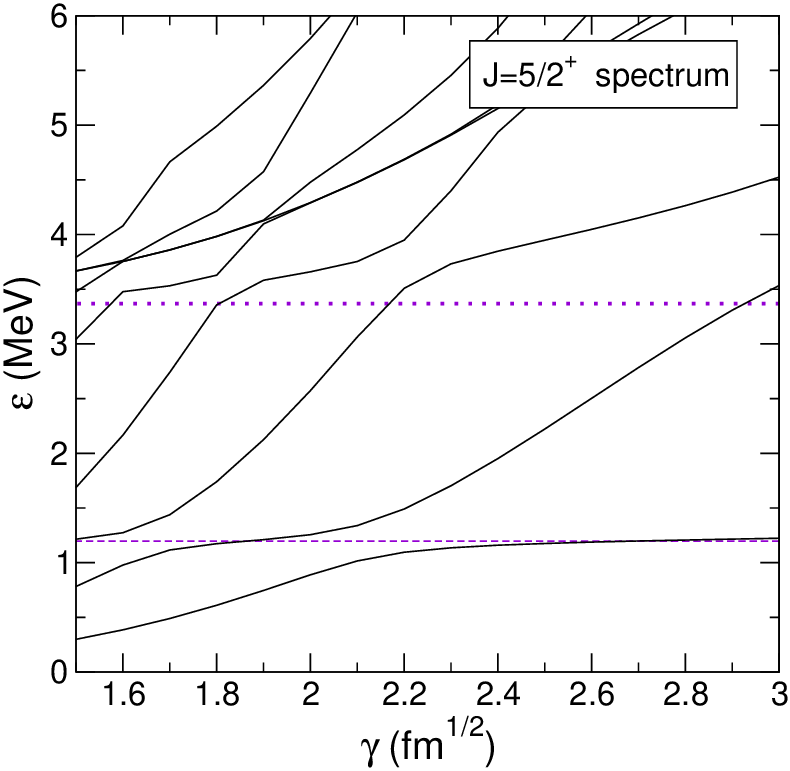}}\par} 
\end{minipage}

\vspace{0.5cm}

\begin{minipage}[b]{0.95\linewidth}
{\par\centering
  \resizebox*{0.95\textwidth}{!}{\includegraphics{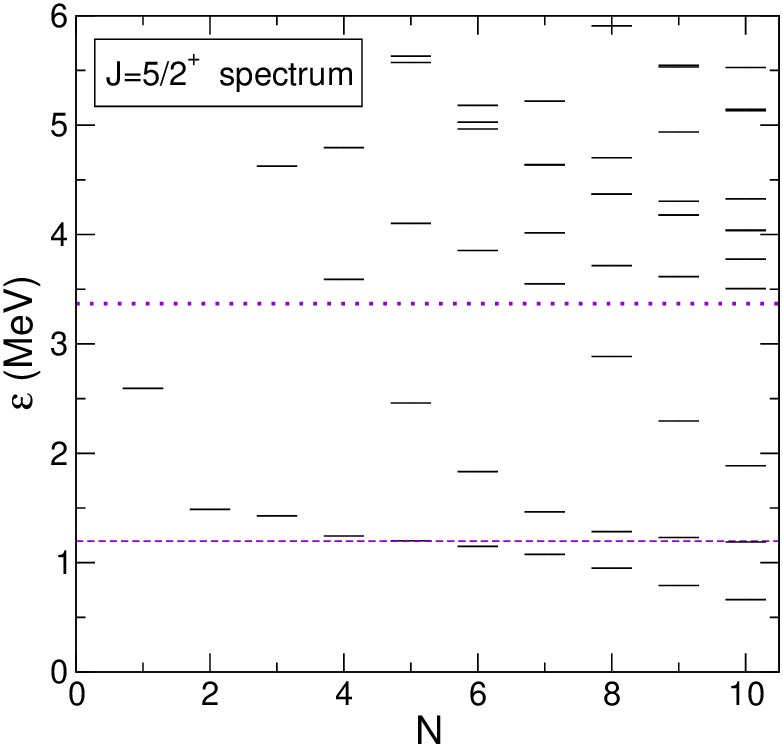}}\par} 
 \caption{\label{be11_j52_spec} (Colour online) Eigenvalues obtained
   from the diagonalization of the  \nuc{11}{Be} Hamiltonian in a THO
   basis, as a function of the LST continuum parameter ($\gamma$) in
   the upper panel, and as a function of the number of oscillator states included in
   the basis in the lower panel. The dashed line indicates the energy of the 5/2$^+$
   resonance and the dotted line, the energy of the $^{10}$Be(2$^+$)+n
   threshold.}  
\end{minipage}
\end{figure}

The selected Hamiltonian contains low-lying resonances at
$\varepsilon=1.2$~MeV ($5/2^+$), 2.7~MeV ($3/2^-$)  
and 3.2~MeV ($3/2^+$) \cite{Nun96}. These values are confirmed
applying the stabilization method  with the THO basis, in the two ways described above.    
As an example, in Fig.~\ref{be11_j52_spec}, we show the results 
 for $J^\pi=5/2^+$. In the upper panel, the sequence of
continuum states with $J^\pi=5/2^+$ is plotted versus the continuum
parameter $\gamma$ of the LST, and for a fixed value of N (N=10). In the lower
panel, the  $J^\pi=5/2^+$ eigenvalues obtained from the
diagonalization of the assumed Hamiltonian in the  THO basis are
plotted as a function of the discrete basis size parameter (N), with $\gamma$ fixed to 1.84~fm$^{1/2}$. The
dashed line marks the known location of the first $5/2^+$ resonance
deduced from the behavior of the phase-shifts and the dotted line marks
the $n$+$^{10}{\rm  Be}(2^+)$ threshold. In both plots,
the  energy stabilization precisely at the nominal energy of the
resonance is apparent.  Similar results are obtained for the  $3/2^+$ and $3/2^-$ 
resonances. 


\begin{figure}
\begin{minipage}[b]{1.0\linewidth}
{\par\centering \resizebox*{0.9\textwidth}{!}{\includegraphics{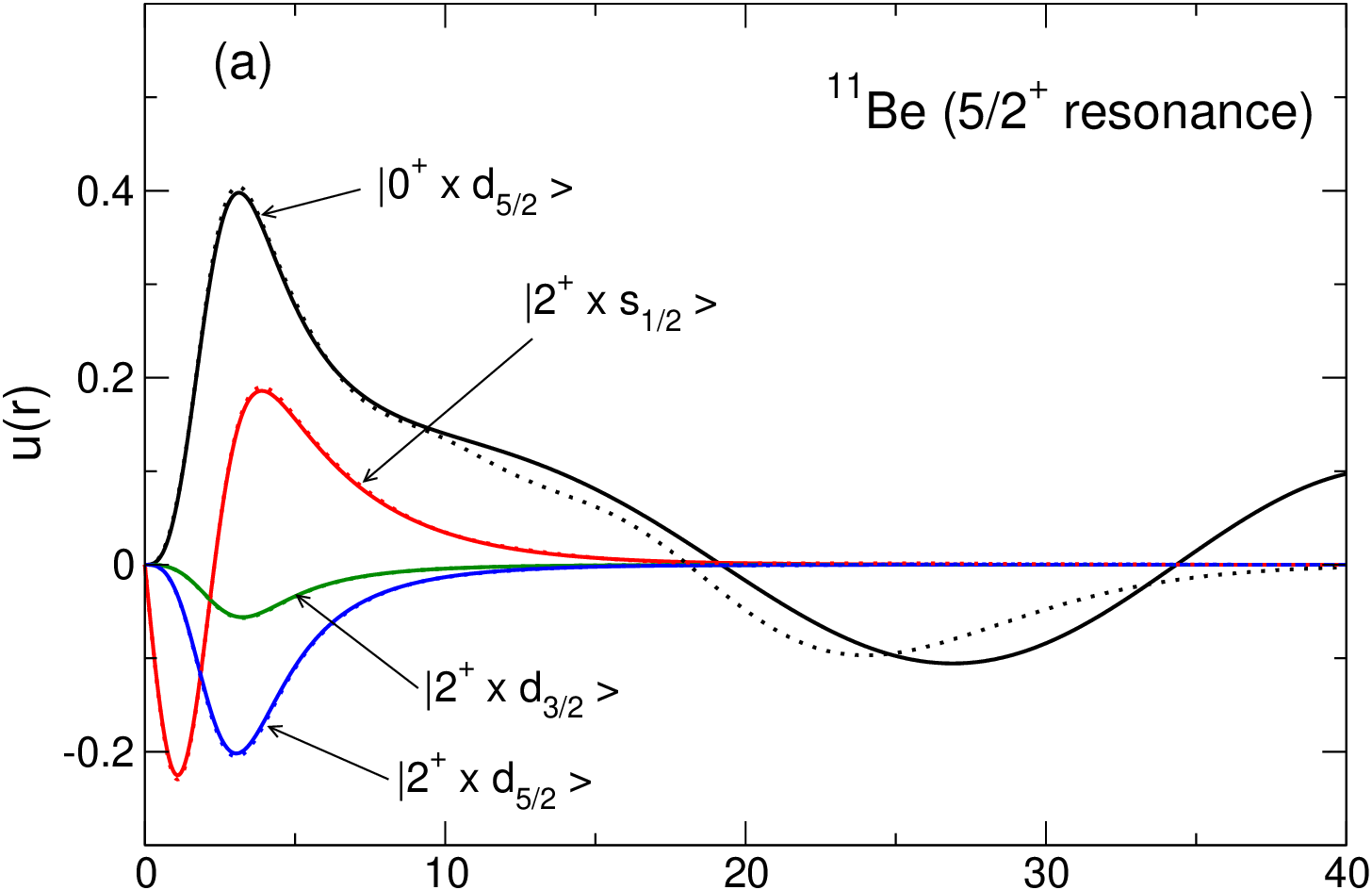}}\par}
\end{minipage}

\vspace{2mm}

\begin{minipage}[b]{1.0\linewidth}
{\par\centering \resizebox*{0.9\textwidth}{!}{\includegraphics{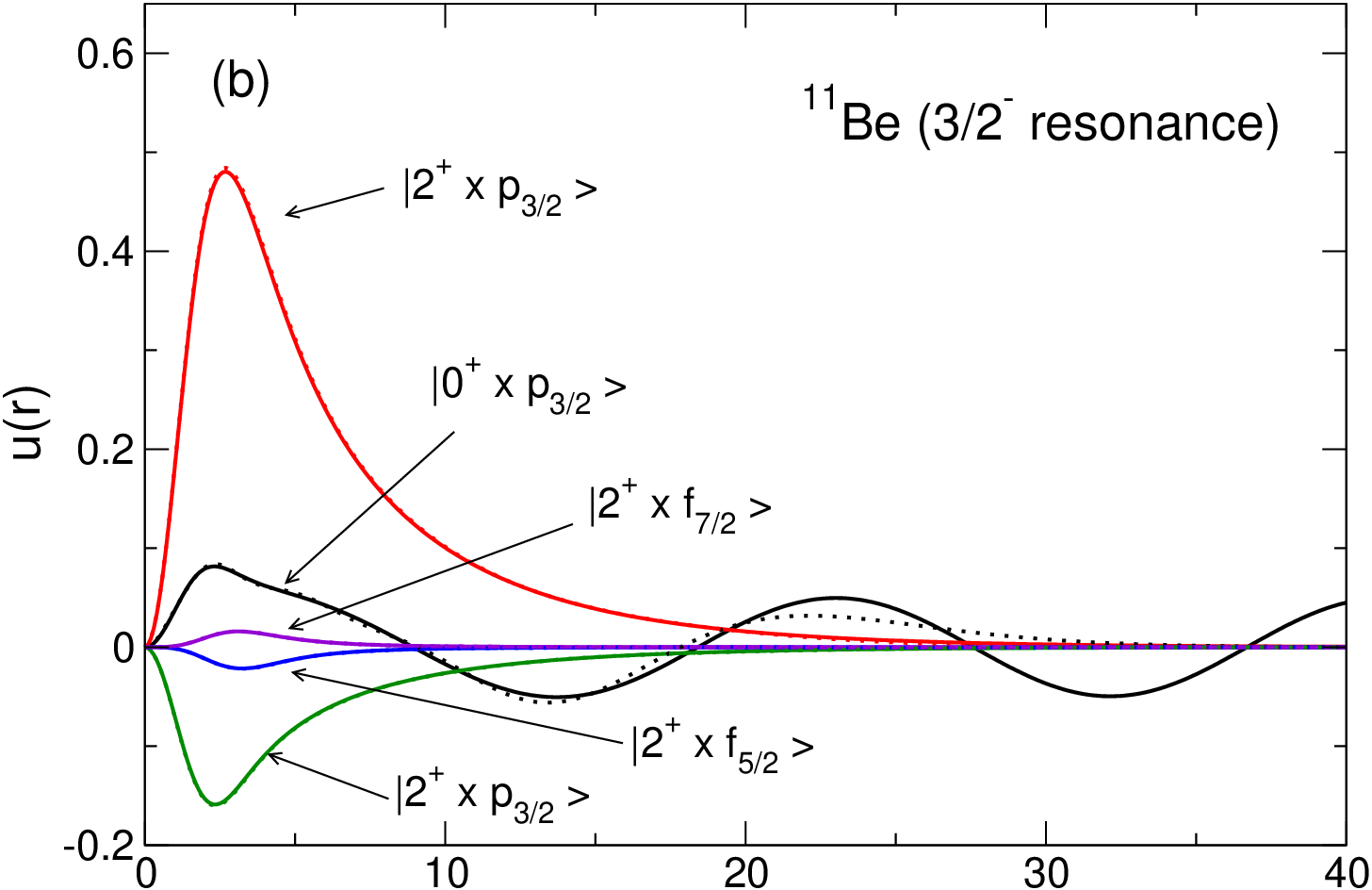}}\par}
\end{minipage}

\vspace{2mm}

\begin{minipage}[b]{1.0\linewidth}
{\par\centering \resizebox*{0.9\textwidth}{!}{\includegraphics{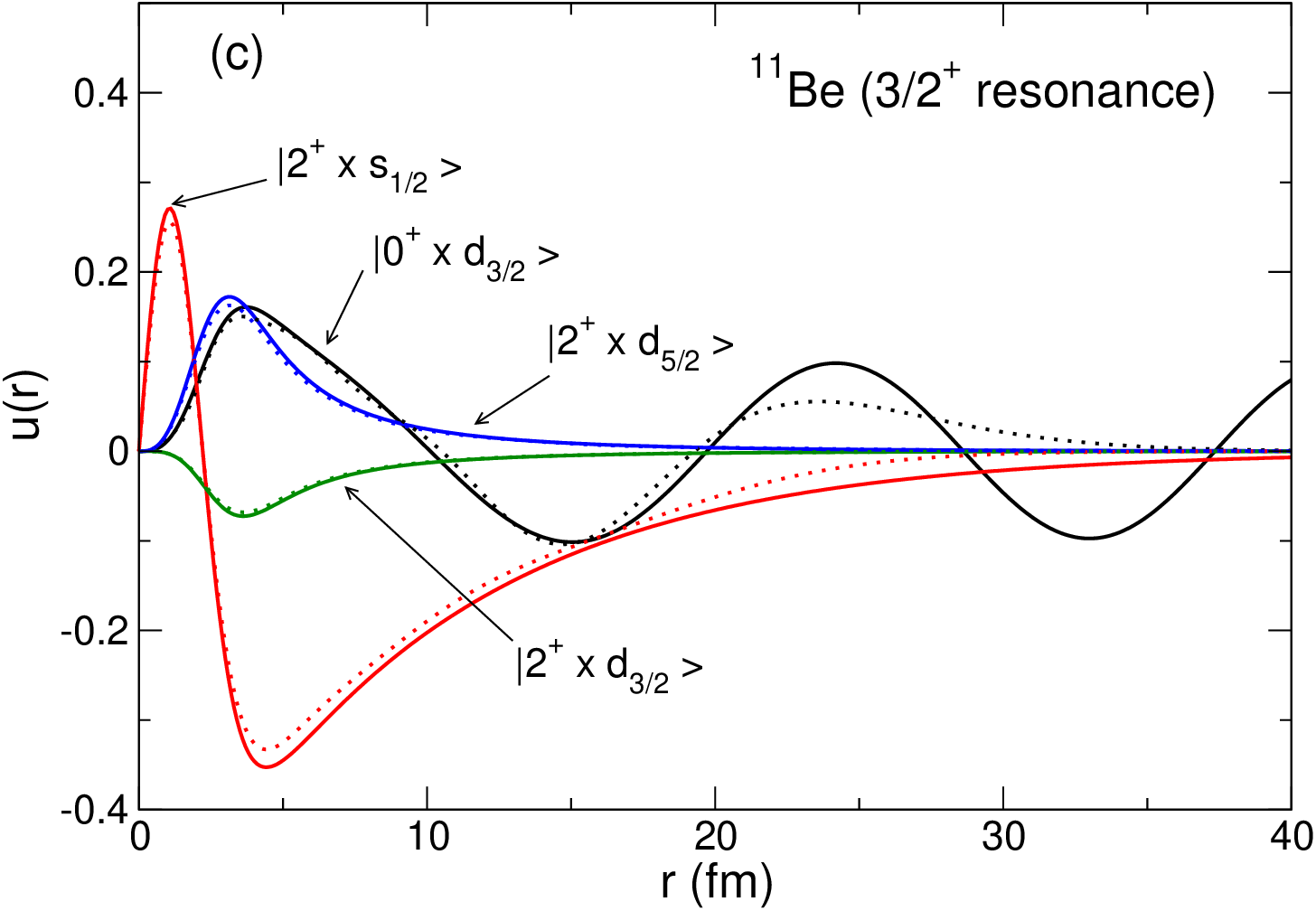}}\par}
\end{minipage}
\caption{\label{wfres} (Colour online) Radial parts of the continuum
   wave functions for the $^{11}$Be resonances at $\varepsilon$=1.2~MeV ($5/2^+$), 
2.7 MeV ($3/2^-$) and 3.2 MeV ($3/2^+$). The solid lines are obtained by direct
   integration of the  Schr\"odinger equation, whereas the dashed
   lines are the result of the diagonalization in a THO basis with
   N=10 ($5/2^+$) or N=9 ($3/2^-$, $3/2^+$) oscillator functions. For a better comparison, the
   normalization of the scattering state has been chosen in order to
   have the same magnitude as the discrete solution at the maximum.} 
\end{figure}

According to the stabilization method, the eigenfunctions corresponding to 
the stabilized energies should   
correspond to well localized states, as expected for a resonant
state. This is confirmed in Fig.~\ref{wfres} for the three resonances discussed above. 
In each panel, we compare  the radial components  of the scattering  
wave functions evaluated at the nominal energy of the resonance (solid lines),
with the THO eigenfunction associated with the stabilized eigenvalue, 
using a basis of N=10  oscillator functions (dashed
lines).  Because the continuum wavefunctions are 
not square-integrable,  these functions have been conveniently 
scaled for a better comparison with the PS functions. 

Note that, for these three resonances, the channels corresponding to $I=2$ are effectively bound,  
since the energy of these resonances is below the $n$+$^{10}{\rm Be}(2^+)$ threshold. The component 
based on the $^{10}{\rm Be}({\rm g.s.})$ is unbound but it shows
the anticipated localization reminiscent of a quasi-stationary
state. We note that, unlike the case of the bound states, we do not expect  
a perfect agreement between both calculations due to the exponential behaviour of the PS basis at 
large distances. 
Apart from that, it is also seen that, in the interior region, the four radial
components are in very good agreement with the exact solution.

\begin{figure}
{\par\centering \resizebox*{0.85\columnwidth}{!}{\includegraphics{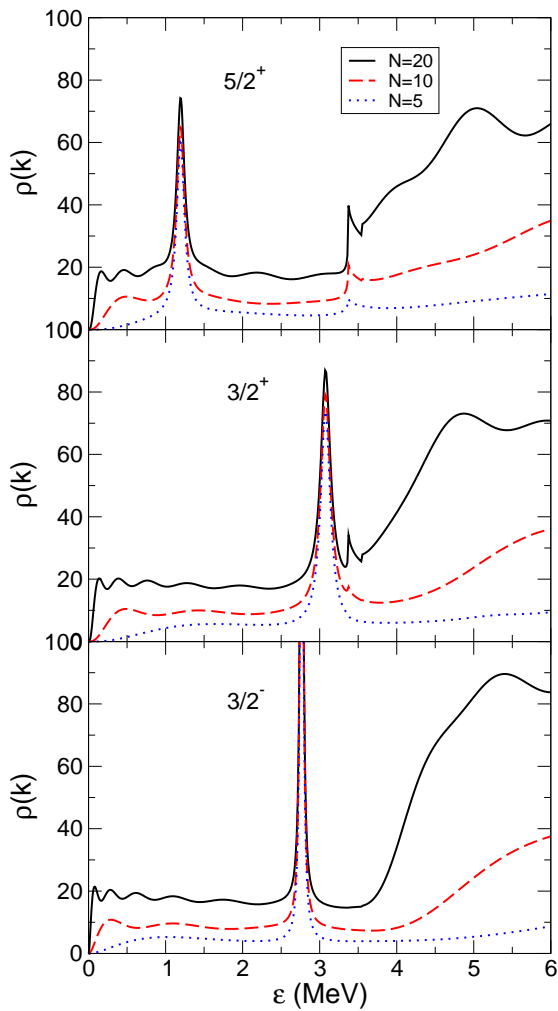}}\par}
 \caption{\label{complet} Density of states for the $5/2^+$,  $3/2^+$  and $3/2^-$ resonances of the $^{11}$Be nucleus, using different 
values of the basis size N.} 
\end{figure}

The stabilization method provides also expressions for the width of
the resonances in the PS basis \cite{Tay76}. However,  these
expressions were originally developed for the single-channel case, and
hence they cannot be directly applied to our case.  To have an
estimate of the width of the resonance we make use the density of states, defined according to Eq.~(\ref{densk}). 
This function is shown in Fig.~\ref{complet} for the discussed resonances, using different values of the basis size (N).  It can be seen  how the density increases as more channels are open above the excitation energy of the core.
It can be seen also that 
the presence of a resonance gives rise to a peak in the density distribution. Based on this property, we have estimated the 
width of the resonance from the FWHM of the corresponding peak in the density distribution. For the $5/2^+$, $3/2^-$ and $3/2^+$ resonances considered above, this method yields $\Gamma=125$~keV, $40$~keV and $140$~keV respectively. This widths are to be compared with the values  reported in \cite{Nun95}, namely, $\Gamma=125$~keV, $50$~keV and 100~keV. Except for the latter, for which our prescription gives a width 
40\% larger, the agreement between both methods is very good in the other two cases.

\begin{figure}
{\par\centering \resizebox*{0.45\textwidth}{!}{\includegraphics{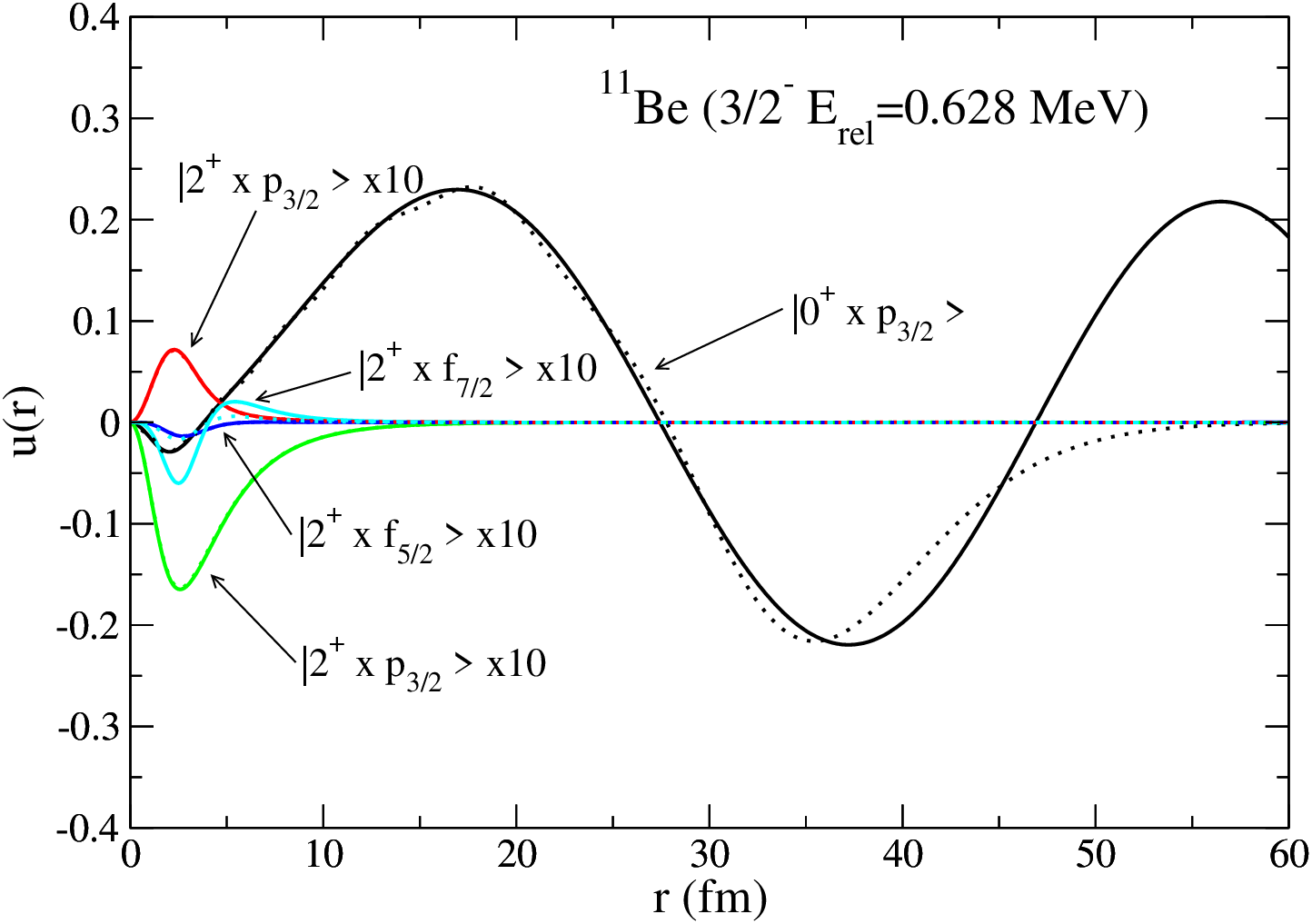}}\par}
 \caption{\label{wf32} (Colour online) Radial parts of the $3/2^{-}$
   wave function for  the \nuc{11}{Be} nucleus at $E_{rel}=0.628$~MeV
   obtained by  direct integration of the Schr\"odinger equation
   (solid lines) and by diagonalization in a THO basis with N=15
   states. All the components except the main one are multiplied by a
   factor of 10.  
}
\end{figure}

Just to complete our study, we show in Fig. \ref{wf32} the comparison
of the radial parts obtained by integration of the Schr\"odinger equation
(solid lines) and by diagonalization in a THO basis with N=15
(dotted lines) for a non-resonant state in the continuum. It can be
observed that the agreement is also good for these states.

\subsection{Electric reduced transitions probabilities}

The electric transition probabilities provide also a useful test to
assess the quality of the basis to represent the continuum
states. These transition probabilities can be calculated using either
the exact scattering states,  using Eq.~(\ref{becontgen}), or the
pseudostates, using Eq.~(\ref{beps}). In the latter case, one obtains
a discrete distribution, which can be converted to a continuous
distribution by means of Eq.~(\ref{befold}). In actual calculations,
this equation is evaluated with a finite number of states (N) and
hence this formula is only approximate. The degree of agreement of
this approximate formula with the exact calculation provides a
measurement of the quality of the PS basis to represent the continuum
for a given operator. In this section we perform this test for the E1
and E2 operators. 

According to Eq.~(\ref{mel}), the electric operator for a valence+core system 
will contain in general contributions coming from the valence excitation, the core 
excitation and mixed excitations. However, 
in our test case, $^{11}$Be, with  core states
restricted to the ground state ($0^+$) and the first excited  
state ($2^+$), dipole transitions will consist of pure
single particle excitations. On the other hand, quadrupole transitions
will contain both single particle and core excitations, but not
simultaneous  transitions. These  simultaneous transitions will only
affect octupole and higher order transitions, which will not be
considered here. 

\begin{figure}
{\par\centering \resizebox*{0.45\textwidth}{!}{\includegraphics{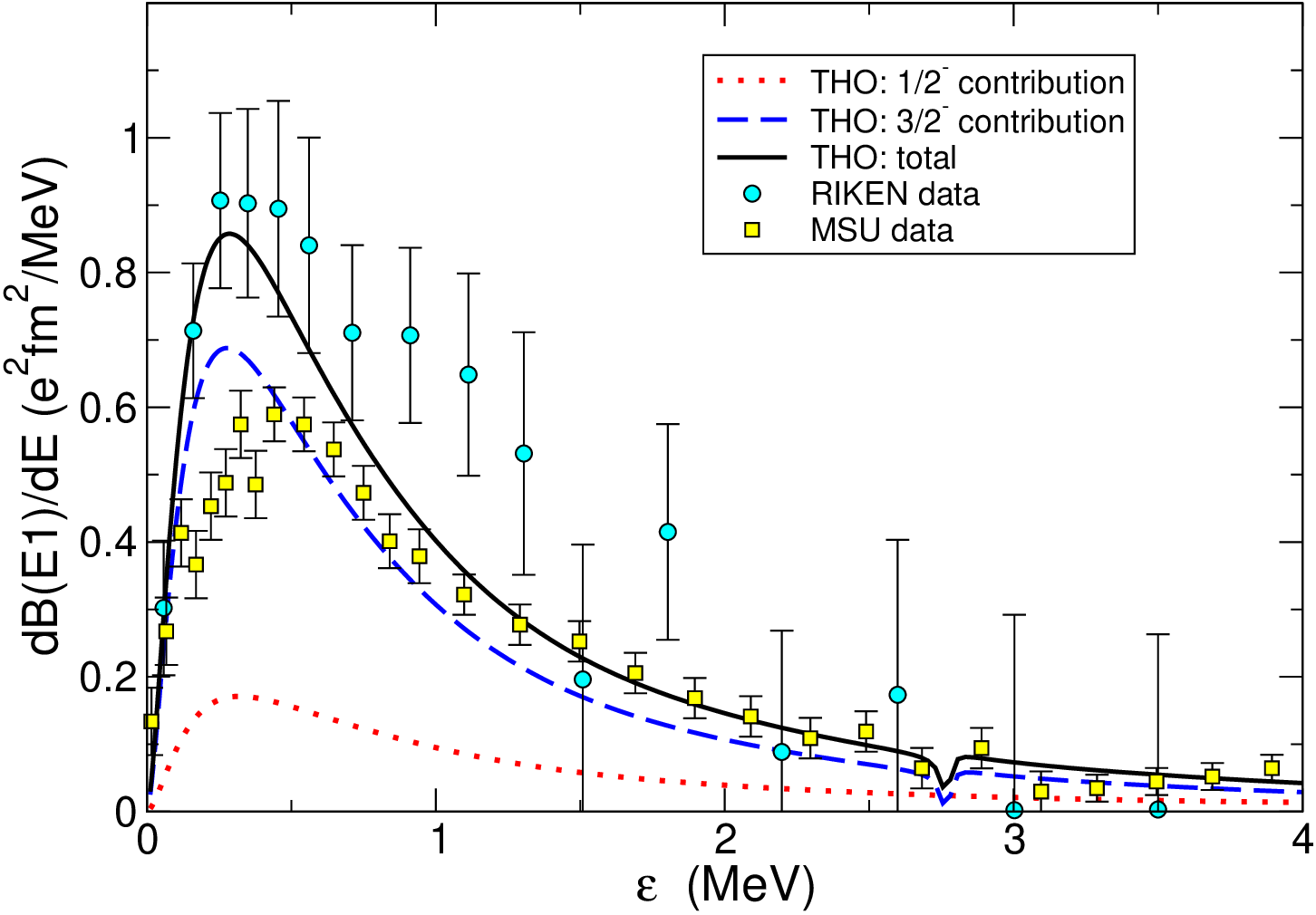}}\par}
 \caption{\label{be1} (Colour online) Dipole electric transition
   probability (${\cal B}(E1)$) obtained with the THO basis and
   compared with experimental data from RIKEN \cite{Nak95} and MSU
   \cite{Pal03}. A THO basis with N=20 was used in the calculation,
   for which the calculated distribution is fully converged and
   indistinguishable from the exact result using the \textit{exact}
   scattering wavefunctions.} 
\end{figure}

In Fig.~\ref{be1}, the energy distribution of the ${\cal B}(E1)$
obtained with a THO basis with N=20 functions is shown for
\nuc{11}{Be}. Separate contributions for  $1/2^-$ and $3/2^-$ states
are shown by dotted and dashed lines, respectively. With this basis
size, the calculated THO distributions are almost indistinguishable
from the exact calculation, obtained with the exact scattering states,
so the latter has not been included in the figure.  
The available experimental distributions from two experiments
performed at RIKEN \cite{Nak95} and MSU \cite{Pal03} are also shown in
the plot. The theoretical distribution lies in between the two
experimental sets of data. However, one has to keep in mind that the RIKEN
data are inclusive with the respect to the $^{10}$Be state and hence
it might contain contributions where the core is left in an excited
state. Moreover, it is also worth noting that the calculation will  
be sensitive to the choice of the $^{11}$Be Hamiltonian. We have not
explored in this work this dependence since the purpose of this
calculation is to test the quality of the basis, rather than a
detailed comparison with the data.  
 
From Fig.~\ref{be1} on sees that the calculated distribution shows a
dip around $\varepsilon = 2.8$~MeV,  which is also visible in the data
from Ref.~\cite{Pal03}. This behavior arises from the presence of the
$3/2^-$ resonance at this excitation energy. This resonance is relatively narrow
($\Gamma = 50$~keV) but it is only weakly coupled because it is mainly
built on the excited core ($^{10}{\rm Be(2^+)}$), whereas the ground
state is mostly $^{10}{\rm Be(0^+)}$.  

Because only the single-particle excitation term of Eq.~(\ref{mel}) contributes to this 
dipole transition, this observable can be also well reproduced 
within a single-particle model of $^{11}$Be, with the $^{10}$Be core
in its ground state, and including the appropriate  spectroscopic
factor for the $| ^{10}{\rm Be}(0^+) \otimes 2s_{1/2}\rangle $ configuration. A
departure from this behaviour is the afore-mentioned reduction of the
${\cal B}(E1)$ around 2.8~MeV which is due to a core dominated
$3/2^{-}$ resonance. 

\begin{figure}
\begin{minipage}[b]{0.95\linewidth}
{\par\centering \resizebox*{0.95\textwidth}{!}{\includegraphics{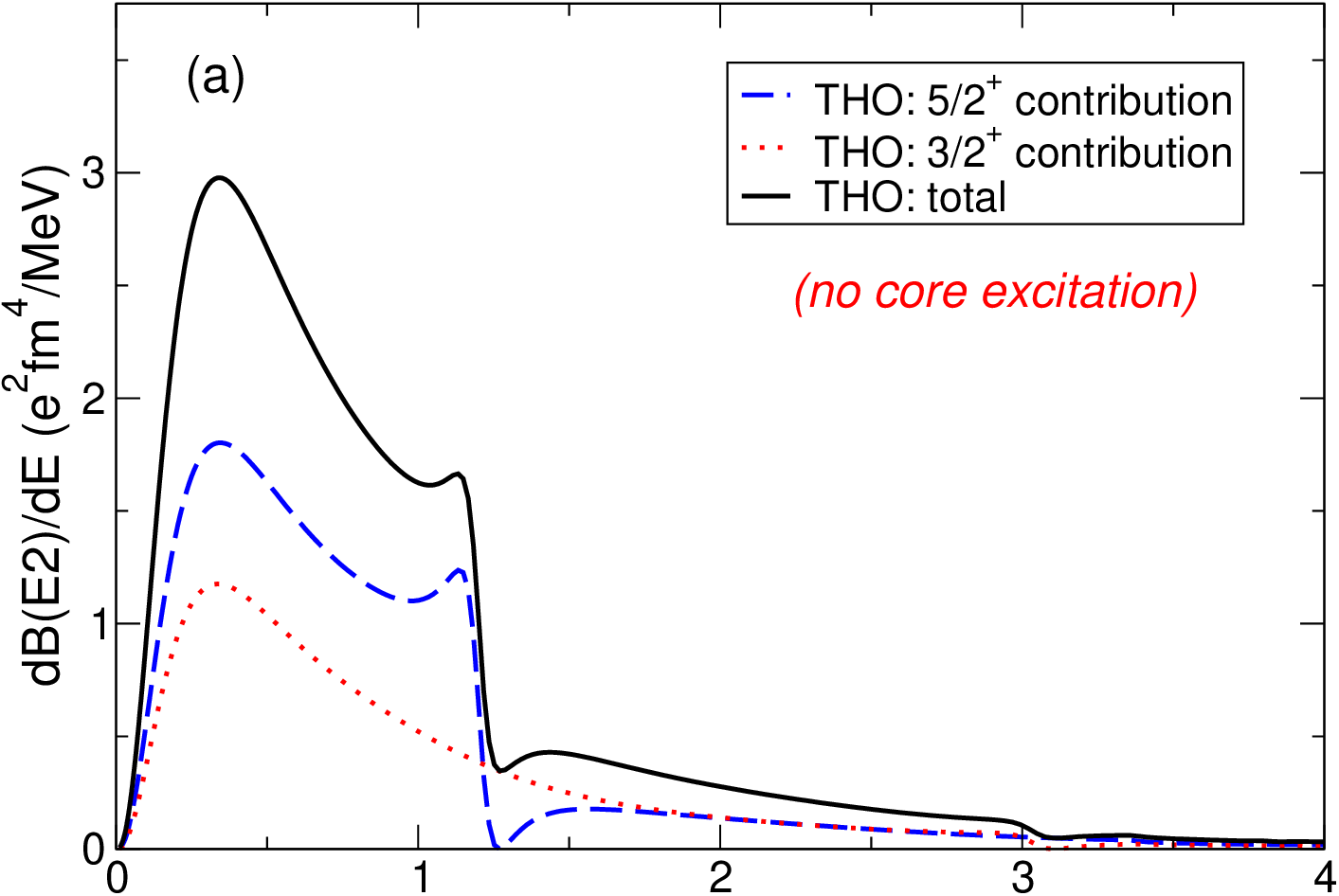}}\par}
\end{minipage}

\vspace{0.5cm}

\begin{minipage}[b]{0.95\linewidth}
{\par\centering \resizebox*{0.95\textwidth}{!}{\includegraphics{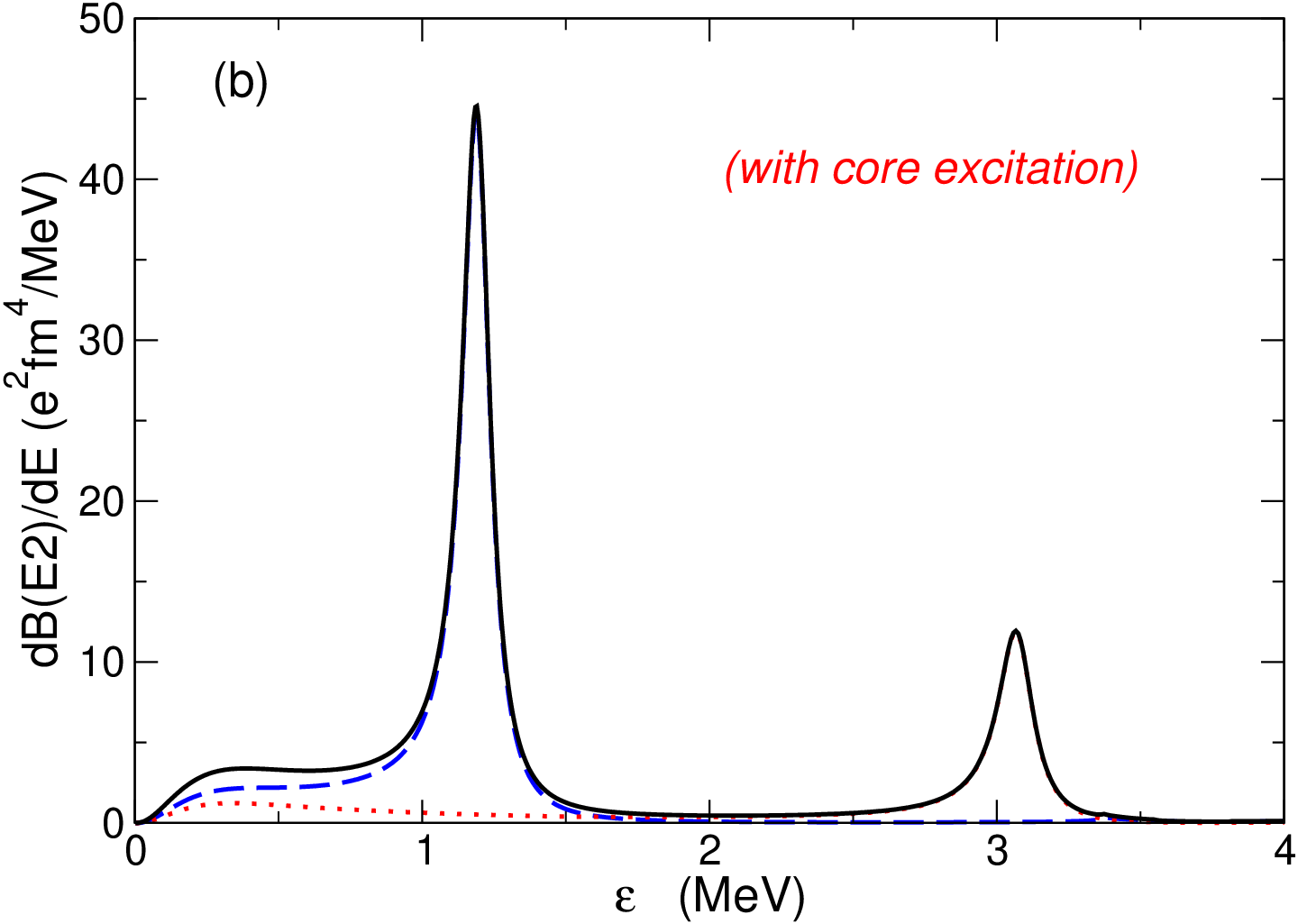}}\par}
\caption{\label{be2} (Colour online) Quadrupole electric transition
  probability (${\cal B}(E2)$)  obtained with the THO basis with
  N=20 oscillator functions. The upper panel is the calculation
  including only the valence excitations. The bottom panel includes
  both valence and core contributions.}  
\end{minipage}
\end{figure}

We have also evaluated the quadrupole electric transition
probabilities, which are shown in Fig.~\ref{be2}. The dotted and
dashed lines are the contributions from $3/2^+$ and $5/2^+$,
respectively, whereas the solid line is the sum of both
contributions. According  to Eq.~(\ref{mel}), in addition to the
single-particle excitations, in this case we have also a contribution
due to E2 transitions of the core which, in  
fact, give the main contribution to the total ${\cal B}(E2)$
strength. To illustrate better the contribution coming from the
valence excitation and the core, we show in the upper panel of this
figure the single-particle contribution, whereas in the bottom panel  
we show the full calculation, including also contributions from the
core. It is seen that  the ${\cal B}(E2)$ strength is dominated by the
core excitations, as expected for a collective transition.  The peaks
at $\varepsilon \simeq 1.2$~MeV and and $\varepsilon \simeq 3.2$ MeV
are due to the $5/2^+$ and $3/2^+$ resonances. Unfortunately, no
experimental or theoretical ${\cal B}(E2)$ for $^{11}$Be has been
found in the literature in order to  compare with.


\subsection{Application to the Coulomb breakup of  $^{11}$Be on $^{208}$Pb}

A more recent measurement of the Coulomb breakup of $^{11}$Be  can be found in  
the work by Fukuda and collaborators \cite{Fuk04},  who measured the breakup
of a $^{11}$Be beam at 69 MeV/nucleon on carbon and lead targets.  

At these energies and for very small angles the breakup is dominated
by the Coulomb interaction. For angles below the grazing angle the
differential break up cross section can be calculated semiclassically
using the equivalent photon method (EPM) \cite{Ber88}. For the E1,
which is expected to be the dominant one, the breakup cross section in
the EPM method reads,  
\begin{equation}
\left(\frac{d^2\sigma}{d\Omega d\varepsilon}\right)_{bu} = \frac{16
  \pi^3}{9 \hbar c} \frac{d{\cal
    B}(E1)}{d\varepsilon}\frac{dN_{E1}(\theta_{cm},E_x)}{d\Omega_{cm}}, 
\label{xsbu1}
\end{equation}
where $N_{E1}(\theta_{cm},E_x)$ denotes the number of virtual photons
with energy $E_x$ at scattering angle $\theta_{cm}$. In this
treatment,  the scattering angle corresponds to a classical Coulomb
trajectory. The photon energy would be always
$E_x=\varepsilon+S_{1n}$. 

\begin{figure}
\begin{minipage}[b]{0.95\linewidth}
{\par\centering
  \resizebox*{0.95\textwidth}{!}{\includegraphics{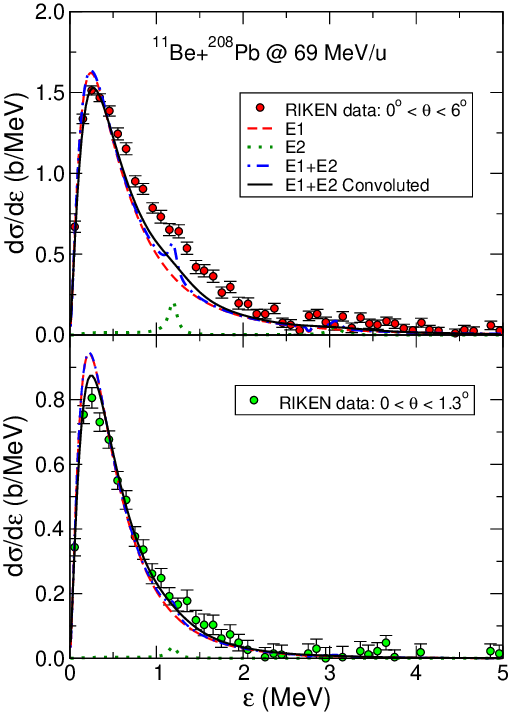}}\par} 
\end{minipage}
\vspace{0.5cm}
\begin{minipage}[b]{0.95\linewidth}
{\par\centering \resizebox*{0.95\textwidth}{!}{\includegraphics{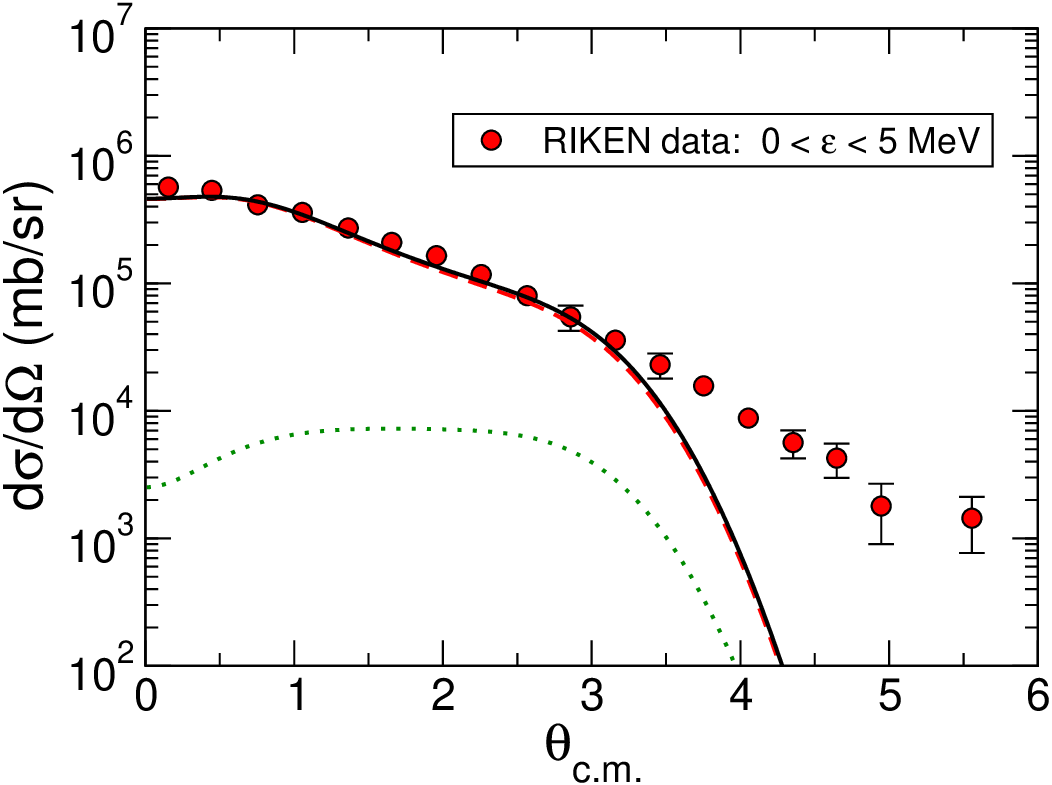}}\par}
 \caption{\label{dsde} (Colour online) Energy (top panel) and angular
   (bottom panel) distributions of the exclusive breakup  for the
   reaction $^{11}$Be+$^{208}$Pb at 69~MeV/nucleon. The circles are
   the experimental data  from Ref.~\cite{Fuk04}. The dot-dashed and dashed
   lines correspond to the E1 and E2 contributions, calculated
   within the  EPM method, and the solid line is the sum of both contributions.}
\end{minipage}
\end{figure}

In a similar way, the E2 contribution to the breakup cross section
which in this formalism is related to the ${\cal B}(E2)$  
distribution,
\begin{equation}
\left(\frac{d^2\sigma}{d\Omega d\varepsilon}\right)_{bu} = \frac{4
  \pi^3}{75 \hbar c} \left( \frac{E_{x}}{\hbar c}\right)^{3}
\frac{d{\cal B}(E2)}{d\varepsilon}\frac{dN_{E2}}{d\Omega_{cm}}.
\label{xsbu2}
\end{equation}
This contribution should be added to the dipole Coulomb break up. The
equivalent photon number for $E2$ transitions can also be found in
\cite{Ber88}. 

We have evaluated these contributions using the ${\cal B}(E1)$  and
${\cal B}(E2)$ distributions obtained with the THO basis. Indeed,
these expressions could be directly evaluated with the scattering
states, since no discretization is required in this case.  It is  
nevertheless illustrative to compare both calculations, to show the
convergence of these observables with the size of the THO basis.  
In more sophisticated reaction models, such as the CDCC method with
core excitation \cite{Summers06},  the use of a discretization method
is mandatory, and hence the use of a discrete basis, like the THO
proposed here, is more justified.  


From the expressions (\ref{xsbu1}) and (\ref{xsbu2}), the energy and
angular differential breakup cross sections are calculated by  
integrating in the scattering angle or in the excitation energy,
respectively. In the former case, a critical ingredient of the  
calculation is the minimum impact parameter ($b_\mathrm{min}$). The
model assumes that pure Coulomb breakup occurs only for
$b>b_\mathrm{min}$. By contrast, for $b<b_\mathrm{min}$, the model
assumes that other reaction mechanisms rather than pure Coulomb
scattering take place (such as nuclear effects). Since these effects
are not properly described by Eqs.~(\ref{xsbu1}) and (\ref{xsbu2}),  
these expressions are only evaluated for $b>b_\mathrm{min}$. 
This is also the reason why these Coulomb breakup experiments are
focused at small angles, ideally below the grazing angle. 
In all the calculations the minimum impact parameter
is settled according to the choice done in \cite{Fuk04},  where the
(E1) Coulomb breakup cross section is also evaluated using the EPM
method.  

In Fig.~\ref{dsde} we compare the calculated energy (upper two-panel
figure) and angular differential cross sections (lower panel) with the
experimental data. The separate E1 and E2 contributions, as well
as their sum, are shown in each panel. The calculations  
have been convoluted with the experimental angular and energy
distributions reported in \cite{Fuk04}.  It is clearly seen that the
main contribution comes from the dipole break up.  In the angular
distribution, the sum of both contributions cannot be distinguished at
the smaller angles from the pure E1 contribution. The small E2 contribution is only observed in the
energy regions of the resonances. This difference is nevertheless  
washed out once the energy resolution of the experiment is
considered. Despite this small contribution, it is observed that  the
E2 component improves the agreement in the energy region nearby the
$5/2^{+}$ resonance. Comparing the total and dipole angular distributions
of the Coulomb break up one can infer up to what angle one should
consider pure E1 excitations in order to extract a ${\cal B}(E1)$
distribution not affected by the 5/2$^{+}$ resonance.

\section{\label{sec:summary} Summary and conclusions}

We have investigated the problem of the description of the states of a
particle moving in a deformed potential in terms of 
a pseudo-state (PS) basis. In the PS method, the states of the system
are approximated by the eigenstates of the Hamiltonian  
in a basis of square-integrable functions. The negative eigenvalues
are identified with the bound states of the system, whereas  
the positive eigenvalues  are regarded as a discrete and finite
representation of the continuum spectrum. Identification of  
resonances is done using the so-called stabilization method
\cite{Haz70,Tay76}. 

Following our previous choice for non-deformed systems, we propose to
use as PS basis the Transformed Harmonic Oscillator (THO)  
basis. The basis functions are obtained by applying an analytic local
scale transformation \cite{Amos,Mor09} to the conventional HO basis.  
The transformation is such that it converts the Gaussian asymptotic
behavior of the HO function into an exponential.  

The method has been applied to the $^{11}$Be nucleus, treated within a
particle-rotor model. The $^{10}$Be core is assumed to  
have a permanent axial deformation with $\beta_2=0.67$
\cite{Nun96}. We have shown that the bound-state energies and wave
functions are very well described using a relatively small basis,
showing perfect agreement with those obtained by direct  
integration of the Schr\"odinger equation. We have shown that the
resonances $5/2^+$, $3/2^-$ and $3/2^+$ are also well described with
the method using small THO bases. It has also been checked that the
wave functions of the non-resonant continuum calculated with the THO
method compare well with the state computed by direct  
integration of the Schr\"odinger equation at the same energy. 

We have given expressions for the   E1 and E2
electric transition probabilities in the discrete basis, and we have
proposed a method to obtain smooth distributions from these discrete
values. To illustrate this method, we have calculated  
the ${\cal B}(E1)$  and ${\cal B}(E2)$  electric transition
probabilities for the $^{11}$Be nucleus. These distributions show a fast
convergence rate with the basis size, and the converged  
results are in perfect agreement with the exact calculation, obtained
with the exact scattering states. With the adopted Hamiltonian,  
the calculated ${\cal B}(E1)$ distribution is consistent, but somewhat
larger, than the experimental data from MSU \cite{Pal03}. 

Finally, we have applied the model to the Coulomb breakup of $^{11}$Be
on $^{208}$Pb at 69~MeV/nucleon, comparing with the data from  
Ref.~\cite{Fuk04}. The reaction is treated in a semi-classical
picture, using  the equivalent photon method, and including both E1
and E2 contributions.   
The calculated angular distribution is in good agreement with the data
for scattering angles below $3^\circ$. Beyond this angle, other  
effects not considered in the EPM method,  such as nuclear breakup,
are expected to take place. The calculated energy distribution is also
in good agreement with the data, particularly when the angular range
is below the grazing angle.  

We conclude that the THO basis provides a suitable representation to describe two-body
composite systems (bound and unbound states) including the core
deformation. This study provides the needed test for accomplishing a
similar study for  more interesting cases, such as three-body composite
systems including core deformation or three-body scattering  problems (two-body projectile plus a target)
including dynamic core  excitation.  Work toward this direction is in progress.

\begin{acknowledgments}
 We are grateful to Ian Thompson for his help in the calculation of the multi-channel  
scattering states.  This work has been partially supported by the Spanish Ministerio de 
 Ciencia e Innovaci\'on and FEDER funds under projects
 FIS2011-28738-c02-01,  FPA2009-07653, 
 FPA2009-08848 and  by the Spanish Consolider-Ingenio 2010 Programme CPAN
(CSD2007-00042)  and by Junta de  Andaluc\'ia (FQM160,
 P07-FQM-02894). J.A.L.\ acknowledges a research grant by the
 Ministerio de Ciencia e Innovaci\'on. 
\end{acknowledgments}

\bibliography{thox-v4}

\begin{thebibliography}{40}
\expandafter\ifx\csname natexlab\endcsname\relax\def\natexlab#1{#1}\fi
\expandafter\ifx\csname bibnamefont\endcsname\relax
  \def\bibnamefont#1{#1}\fi
\expandafter\ifx\csname bibfnamefont\endcsname\relax
  \def\bibfnamefont#1{#1}\fi
\expandafter\ifx\csname citenamefont\endcsname\relax
  \def\citenamefont#1{#1}\fi
\expandafter\ifx\csname url\endcsname\relax
  \def\url#1{\texttt{#1}}\fi
\expandafter\ifx\csname urlprefix\endcsname\relax\def\urlprefix{URL }\fi
\providecommand{\bibinfo}[2]{#2}
\providecommand{\eprint}[2][]{\url{#2}}

\bibitem[{\citenamefont{Rawitscher}(1974)}]{Raw74}
\bibinfo{author}{\bibfnamefont{G.~H.} \bibnamefont{Rawitscher}},
  \bibinfo{journal}{Phys. Rev. C} \textbf{\bibinfo{volume}{9}},
  \bibinfo{pages}{2210} (\bibinfo{year}{1974}).

\bibitem[{\citenamefont{Austern et~al.}(1987)\citenamefont{Austern, Iseri,
  Kamimura, Kawai, Rawitscher, and Yahiro}}]{Aus87}
\bibinfo{author}{\bibfnamefont{N.}~\bibnamefont{Austern}},
  \bibinfo{author}{\bibfnamefont{Y.}~\bibnamefont{Iseri}},
  \bibinfo{author}{\bibfnamefont{M.}~\bibnamefont{Kamimura}},
  \bibinfo{author}{\bibfnamefont{M.}~\bibnamefont{Kawai}},
  \bibinfo{author}{\bibfnamefont{G.}~\bibnamefont{Rawitscher}},
  \bibnamefont{and} \bibinfo{author}{\bibfnamefont{M.}~\bibnamefont{Yahiro}},
  \bibinfo{journal}{Phys. Rep.} \textbf{\bibinfo{volume}{154}},
  \bibinfo{pages}{125} (\bibinfo{year}{1987}).

\bibitem[{\citenamefont{{Summers} et~al.}(2006)\citenamefont{{Summers},
  {Nunes}, and {Thompson}}}]{Summers06}
\bibinfo{author}{\bibfnamefont{N.~C.} \bibnamefont{{Summers}}},
  \bibinfo{author}{\bibfnamefont{F.~M.} \bibnamefont{{Nunes}}},
  \bibnamefont{and} \bibinfo{author}{\bibfnamefont{I.~J.}
  \bibnamefont{{Thompson}}}, \bibinfo{journal}{Phys. Rev. C}
  \textbf{\bibinfo{volume}{74}}, \bibinfo{pages}{014606}
  (\bibinfo{year}{2006}).

\bibitem[{\citenamefont{Bohr and Mottelson}(1969)}]{BM}
\bibinfo{author}{\bibfnamefont{A.}~\bibnamefont{Bohr}} \bibnamefont{and}
  \bibinfo{author}{\bibfnamefont{B.}~\bibnamefont{Mottelson}},
  \emph{\bibinfo{title}{Nuclear Structure}} (\bibinfo{year}{1969}),
  \bibinfo{edition}{{New York, W. A. Benjamin}} ed.

\bibitem[{\citenamefont{{Matsumoto} et~al.}(2003)\citenamefont{{Matsumoto},
  {Kamizato}, {Ogata}, {Iseri}, {Hiyama}, {Kamimura}, and {Yahiro}}}]{Mat03}
\bibinfo{author}{\bibfnamefont{T.}~\bibnamefont{{Matsumoto}}},
  \bibinfo{author}{\bibfnamefont{T.}~\bibnamefont{{Kamizato}}},
  \bibinfo{author}{\bibfnamefont{K.}~\bibnamefont{{Ogata}}},
  \bibinfo{author}{\bibfnamefont{Y.}~\bibnamefont{{Iseri}}},
  \bibinfo{author}{\bibfnamefont{E.}~\bibnamefont{{Hiyama}}},
  \bibinfo{author}{\bibfnamefont{M.}~\bibnamefont{{Kamimura}}},
  \bibnamefont{and} \bibinfo{author}{\bibfnamefont{M.}~\bibnamefont{{Yahiro}}},
  \bibinfo{journal}{Phys. Rev. C} \textbf{\bibinfo{volume}{68}},
  \bibinfo{pages}{064607} (\bibinfo{year}{2003}).

\bibitem[{\citenamefont{P{\'e}rez-Bernal
  et~al.}(2002)\citenamefont{P{\'e}rez-Bernal, Martel, Arias, and
  G{\'o}mez-Camacho}}]{Per02}
\bibinfo{author}{\bibfnamefont{F.}~\bibnamefont{P{\'e}rez-Bernal}},
  \bibinfo{author}{\bibfnamefont{I.}~\bibnamefont{Martel}},
  \bibinfo{author}{\bibfnamefont{J.~M.} \bibnamefont{Arias}}, \bibnamefont{and}
  \bibinfo{author}{\bibfnamefont{J.}~\bibnamefont{G{\'o}mez-Camacho}},
  \bibinfo{journal}{Few-Body Syst. Suppl.} \textbf{\bibinfo{volume}{13}},
  \bibinfo{pages}{217} (\bibinfo{year}{2002}).

\bibitem[{\citenamefont{Moro et~al.}(2009)\citenamefont{Moro, Arias,
  G{\'o}mez-Camacho, and P{\'e}rez-Bernal}}]{Mor09}
\bibinfo{author}{\bibfnamefont{A.~M.} \bibnamefont{Moro}},
  \bibinfo{author}{\bibfnamefont{J.~M.} \bibnamefont{Arias}},
  \bibinfo{author}{\bibfnamefont{J.}~\bibnamefont{G{\'o}mez-Camacho}},
  \bibnamefont{and}
  \bibinfo{author}{\bibfnamefont{F.}~\bibnamefont{P{\'e}rez-Bernal}},
  \bibinfo{journal}{Phys. Rev. C} \textbf{\bibinfo{volume}{80}},
  \bibinfo{pages}{054605} (\bibinfo{year}{2009}).

\bibitem[{\citenamefont{{Matsumoto}
  et~al.}(2004{\natexlab{a}})\citenamefont{{Matsumoto}, {Hiyama}, {Yahiro},
  {Ogata}, {Iseri}, and {Kamimura}}}]{Mat04a}
\bibinfo{author}{\bibfnamefont{T.}~\bibnamefont{{Matsumoto}}},
  \bibinfo{author}{\bibfnamefont{E.}~\bibnamefont{{Hiyama}}},
  \bibinfo{author}{\bibfnamefont{M.}~\bibnamefont{{Yahiro}}},
  \bibinfo{author}{\bibfnamefont{K.}~\bibnamefont{{Ogata}}},
  \bibinfo{author}{\bibfnamefont{Y.}~\bibnamefont{{Iseri}}}, \bibnamefont{and}
  \bibinfo{author}{\bibfnamefont{M.}~\bibnamefont{{Kamimura}}},
  \bibinfo{journal}{Nucl. Phys. A} \textbf{\bibinfo{volume}{738}},
  \bibinfo{pages}{471} (\bibinfo{year}{2004}{\natexlab{a}}).

\bibitem[{\citenamefont{{Matsumoto}
  et~al.}(2004{\natexlab{b}})\citenamefont{{Matsumoto}, {Hiyama}, {Ogata},
  {Iseri}, {Kamimura}, {Chiba}, and {Yahiro}}}]{Mat04b}
\bibinfo{author}{\bibfnamefont{T.}~\bibnamefont{{Matsumoto}}},
  \bibinfo{author}{\bibfnamefont{E.}~\bibnamefont{{Hiyama}}},
  \bibinfo{author}{\bibfnamefont{K.}~\bibnamefont{{Ogata}}},
  \bibinfo{author}{\bibfnamefont{Y.}~\bibnamefont{{Iseri}}},
  \bibinfo{author}{\bibfnamefont{M.}~\bibnamefont{{Kamimura}}},
  \bibinfo{author}{\bibfnamefont{S.}~\bibnamefont{{Chiba}}}, \bibnamefont{and}
  \bibinfo{author}{\bibfnamefont{M.}~\bibnamefont{{Yahiro}}},
  \bibinfo{journal}{Phys. Rev. C} \textbf{\bibinfo{volume}{70}},
  \bibinfo{pages}{061601(R)} (\bibinfo{year}{2004}{\natexlab{b}}).

\bibitem[{\citenamefont{{Matsumoto} et~al.}(2006)\citenamefont{{Matsumoto},
  {Egami}, {Ogata}, {Iseri}, {Kamimura}, and {Yahiro}}}]{Mat06}
\bibinfo{author}{\bibfnamefont{T.}~\bibnamefont{{Matsumoto}}},
  \bibinfo{author}{\bibfnamefont{T.}~\bibnamefont{{Egami}}},
  \bibinfo{author}{\bibfnamefont{K.}~\bibnamefont{{Ogata}}},
  \bibinfo{author}{\bibfnamefont{Y.}~\bibnamefont{{Iseri}}},
  \bibinfo{author}{\bibfnamefont{M.}~\bibnamefont{{Kamimura}}},
  \bibnamefont{and} \bibinfo{author}{\bibfnamefont{M.}~\bibnamefont{{Yahiro}}},
  \bibinfo{journal}{Phys. Rev. C} \textbf{\bibinfo{volume}{73}},
  \bibinfo{pages}{051602(R)} (\bibinfo{year}{2006}).

\bibitem[{\citenamefont{Rodr{\'i}guez-Gallardo
  et~al.}(2008)\citenamefont{Rodr{\'i}guez-Gallardo, Arias, G{\'o}mez-Camacho,
  Johnson, Moro, Thompson, and Tostevin}}]{manoli08}
\bibinfo{author}{\bibfnamefont{M.}~\bibnamefont{Rodr{\'i}guez-Gallardo}},
  \bibinfo{author}{\bibfnamefont{J.~M.} \bibnamefont{Arias}},
  \bibinfo{author}{\bibfnamefont{J.}~\bibnamefont{G{\'o}mez-Camacho}},
  \bibinfo{author}{\bibfnamefont{R.~C.} \bibnamefont{Johnson}},
  \bibinfo{author}{\bibfnamefont{A.~M.} \bibnamefont{Moro}},
  \bibinfo{author}{\bibfnamefont{I.~J.} \bibnamefont{Thompson}},
  \bibnamefont{and} \bibinfo{author}{\bibfnamefont{J.~A.}
  \bibnamefont{Tostevin}}, \bibinfo{journal}{Phys. Rev. C}
  \textbf{\bibinfo{volume}{77}}, \bibinfo{pages}{064609}
  (\bibinfo{year}{2008}).

\bibitem[{\citenamefont{{Vautherin}}(1973)}]{Vau73}
\bibinfo{author}{\bibfnamefont{D.}~\bibnamefont{{Vautherin}}},
  \bibinfo{journal}{Phys. Rev. C} \textbf{\bibinfo{volume}{7}},
  \bibinfo{pages}{296} (\bibinfo{year}{1973}).

\bibitem[{\citenamefont{{Gambhir} et~al.}(1990)\citenamefont{{Gambhir}, {Ring},
  and {Thimet}}}]{Gam90}
\bibinfo{author}{\bibfnamefont{Y.~K.} \bibnamefont{{Gambhir}}},
  \bibinfo{author}{\bibfnamefont{P.}~\bibnamefont{{Ring}}}, \bibnamefont{and}
  \bibinfo{author}{\bibfnamefont{A.}~\bibnamefont{{Thimet}}},
  \bibinfo{journal}{Ann. Phys. (New York)} \textbf{\bibinfo{volume}{198}},
  \bibinfo{pages}{132} (\bibinfo{year}{1990}).

\bibitem[{\citenamefont{{Zhou} et~al.}(2003)\citenamefont{{Zhou}, {Meng}, and
  {Ring}}}]{Zho03}
\bibinfo{author}{\bibfnamefont{S.~G.} \bibnamefont{{Zhou}}},
  \bibinfo{author}{\bibfnamefont{J.}~\bibnamefont{{Meng}}}, \bibnamefont{and}
  \bibinfo{author}{\bibfnamefont{P.}~\bibnamefont{{Ring}}},
  \bibinfo{journal}{Phys. Rev. C} \textbf{\bibinfo{volume}{68}},
  \bibinfo{pages}{034323} (\bibinfo{year}{2003}).

\bibitem[{\citenamefont{{Bang} and {Vaagen}}(1980)}]{Ban80}
\bibinfo{author}{\bibfnamefont{J.~M.} \bibnamefont{{Bang}}} \bibnamefont{and}
  \bibinfo{author}{\bibfnamefont{J.~S.} \bibnamefont{{Vaagen}}},
  \bibinfo{journal}{Z. Phys. A} \textbf{\bibinfo{volume}{297}},
  \bibinfo{pages}{223} (\bibinfo{year}{1980}).

\bibitem[{\citenamefont{Nunes et~al.}(1996)\citenamefont{Nunes, Christley,
  Thompson, Johnson, and Efros}}]{Nun96}
\bibinfo{author}{\bibfnamefont{F.}~\bibnamefont{Nunes}},
  \bibinfo{author}{\bibfnamefont{J.}~\bibnamefont{Christley}},
  \bibinfo{author}{\bibfnamefont{I.}~\bibnamefont{Thompson}},
  \bibinfo{author}{\bibfnamefont{R.}~\bibnamefont{Johnson}}, \bibnamefont{and}
  \bibinfo{author}{\bibfnamefont{V.}~\bibnamefont{Efros}},
  \bibinfo{journal}{Nucl. Phys. A} \textbf{\bibinfo{volume}{609}},
  \bibinfo{pages}{43} (\bibinfo{year}{1996}).

\bibitem[{\citenamefont{{Karataglidis}
  et~al.}(2005)\citenamefont{{Karataglidis}, {Amos}, and {Giraud}}}]{Amos}
\bibinfo{author}{\bibfnamefont{S.}~\bibnamefont{{Karataglidis}}},
  \bibinfo{author}{\bibfnamefont{K.}~\bibnamefont{{Amos}}}, \bibnamefont{and}
  \bibinfo{author}{\bibfnamefont{B.~G.} \bibnamefont{{Giraud}}},
  \bibinfo{journal}{Phys. Rev. C} \textbf{\bibinfo{volume}{71}},
  \bibinfo{pages}{064601} (\bibinfo{year}{2005}).

\bibitem[{\citenamefont{{Lay} et~al.}(2010{\natexlab{a}})\citenamefont{{Lay},
  {Moro}, {Arias}, and {Gomez-Camacho}}}]{Lay10}
\bibinfo{author}{\bibfnamefont{J.~A.} \bibnamefont{{Lay}}},
  \bibinfo{author}{\bibfnamefont{A.~M.} \bibnamefont{{Moro}}},
  \bibinfo{author}{\bibfnamefont{J.~M.} \bibnamefont{{Arias}}},
  \bibnamefont{and}
  \bibinfo{author}{\bibfnamefont{J.}~\bibnamefont{{Gomez-Camacho}}},
  \bibinfo{journal}{Phys. Rev. C} \textbf{\bibinfo{volume}{82}},
  \bibinfo{pages}{024605} (\bibinfo{year}{2010}{\natexlab{a}}).

\bibitem[{\citenamefont{Hazi and Taylor}(1970)}]{Haz70}
\bibinfo{author}{\bibfnamefont{A.~U.} \bibnamefont{Hazi}} \bibnamefont{and}
  \bibinfo{author}{\bibfnamefont{H.~S.} \bibnamefont{Taylor}},
  \bibinfo{journal}{Phys. Rev. A} \textbf{\bibinfo{volume}{1}},
  \bibinfo{pages}{1109} (\bibinfo{year}{1970}).

\bibitem[{\citenamefont{Taylor and Hazi}(1976)}]{Tay76}
\bibinfo{author}{\bibfnamefont{H.~S.} \bibnamefont{Taylor}} \bibnamefont{and}
  \bibinfo{author}{\bibfnamefont{A.~U.} \bibnamefont{Hazi}},
  \bibinfo{journal}{Phys. Rev. A} \textbf{\bibinfo{volume}{14}},
  \bibinfo{pages}{2071} (\bibinfo{year}{1976}).

\bibitem[{\citenamefont{Stoitsov and Petkov}(1988)}]{SP88}
\bibinfo{author}{\bibfnamefont{M.~V.} \bibnamefont{Stoitsov}} \bibnamefont{and}
  \bibinfo{author}{\bibfnamefont{I.~Z.} \bibnamefont{Petkov}},
  \bibinfo{journal}{Ann. Phys. (N. Y.)} \textbf{\bibinfo{volume}{{\bf 184}}},
  \bibinfo{pages}{121} (\bibinfo{year}{1988}).

\bibitem[{\citenamefont{Petkov and Stoitsov}(1991)}]{PS91}
\bibinfo{author}{\bibfnamefont{I.~Z.} \bibnamefont{Petkov}} \bibnamefont{and}
  \bibinfo{author}{\bibfnamefont{M.~V.} \bibnamefont{Stoitsov}},
  \emph{\bibinfo{title}{Nuclear Density Functional Theory, Oxford Studies in
  Physics}} (\bibinfo{publisher}{Clarendon, Oxford}, \bibinfo{year}{1991}).

\bibitem[{\citenamefont{P{\'e}rez-Bernal
  et~al.}(2001)\citenamefont{P{\'e}rez-Bernal, Martel, Arias, and
  G{\'o}mez-Camacho}}]{Per01}
\bibinfo{author}{\bibfnamefont{F.}~\bibnamefont{P{\'e}rez-Bernal}},
  \bibinfo{author}{\bibfnamefont{I.}~\bibnamefont{Martel}},
  \bibinfo{author}{\bibfnamefont{J.~M.} \bibnamefont{Arias}}, \bibnamefont{and}
  \bibinfo{author}{\bibfnamefont{J.}~\bibnamefont{G{\'o}mez-Camacho}},
  \bibinfo{journal}{Phys. Rev. A} \textbf{\bibinfo{volume}{63}},
  \bibinfo{pages}{052111} (\bibinfo{year}{2001}).

\bibitem[{\citenamefont{Tostevin et~al.}(2001)\citenamefont{Tostevin, Nunes,
  and Thompson}}]{Tos01}
\bibinfo{author}{\bibfnamefont{J.~A.} \bibnamefont{Tostevin}},
  \bibinfo{author}{\bibfnamefont{F.~M.} \bibnamefont{Nunes}}, \bibnamefont{and}
  \bibinfo{author}{\bibfnamefont{I.~J.} \bibnamefont{Thompson}},
  \bibinfo{journal}{Phys. Rev. C} \textbf{\bibinfo{volume}{63}},
  \bibinfo{pages}{024617} (\bibinfo{year}{2001}).

\bibitem[{\citenamefont{Brink and {Satchler}}(1968)}]{BS}
\bibinfo{author}{\bibfnamefont{D.~M.} \bibnamefont{Brink}} \bibnamefont{and}
  \bibinfo{author}{\bibfnamefont{G.~R.} \bibnamefont{{Satchler}}},
  \emph{\bibinfo{title}{Angular Momentum}} (\bibinfo{publisher}{Clarendon,
  Oxford}, \bibinfo{year}{1968}).

\bibitem[{\citenamefont{{Lay} et~al.}(2010{\natexlab{b}})\citenamefont{{Lay},
  {Fedorov}, {Jensen}, {Garrido}, and {Romero-Redondo}}}]{lay10a}
\bibinfo{author}{\bibfnamefont{J.~A.} \bibnamefont{{Lay}}},
  \bibinfo{author}{\bibfnamefont{D.~V.} \bibnamefont{{Fedorov}}},
  \bibinfo{author}{\bibfnamefont{A.~S.} \bibnamefont{{Jensen}}},
  \bibinfo{author}{\bibfnamefont{E.}~\bibnamefont{{Garrido}}},
  \bibnamefont{and}
  \bibinfo{author}{\bibfnamefont{C.}~\bibnamefont{{Romero-Redondo}}},
  \bibinfo{journal}{Eur. Phys. Jour. A} \textbf{\bibinfo{volume}{44}},
  \bibinfo{pages}{261} (\bibinfo{year}{2010}{\natexlab{b}}).

\bibitem[{\citenamefont{{Typel} and {Baur}}(2005)}]{Typ05}
\bibinfo{author}{\bibfnamefont{S.}~\bibnamefont{{Typel}}} \bibnamefont{and}
  \bibinfo{author}{\bibfnamefont{G.}~\bibnamefont{{Baur}}},
  \bibinfo{journal}{Nucl. Phys. A} \textbf{\bibinfo{volume}{759}},
  \bibinfo{pages}{247} (\bibinfo{year}{2005}).

\bibitem[{\citenamefont{Moro et~al.}(2006)\citenamefont{Moro, P{\'e}rez-Bernal,
  Arias, and G{\'o}mez-Camacho}}]{Mor06}
\bibinfo{author}{\bibfnamefont{A.~M.} \bibnamefont{Moro}},
  \bibinfo{author}{\bibfnamefont{F.}~\bibnamefont{P{\'e}rez-Bernal}},
  \bibinfo{author}{\bibfnamefont{J.~M.} \bibnamefont{Arias}}, \bibnamefont{and}
  \bibinfo{author}{\bibfnamefont{J.}~\bibnamefont{G{\'o}mez-Camacho}},
  \bibinfo{journal}{Phys. Rev. C} \textbf{\bibinfo{volume}{73}},
  \bibinfo{pages}{044612} (\bibinfo{year}{2006}).

\bibitem[{\citenamefont{Capel et~al.}(2004)\citenamefont{Capel, Goldstein, and
  Baye}}]{Cap04}
\bibinfo{author}{\bibfnamefont{P.}~\bibnamefont{Capel}},
  \bibinfo{author}{\bibfnamefont{G.}~\bibnamefont{Goldstein}},
  \bibnamefont{and} \bibinfo{author}{\bibfnamefont{D.}~\bibnamefont{Baye}},
  \bibinfo{journal}{Phys. Rev. C} \textbf{\bibinfo{volume}{70}},
  \bibinfo{pages}{064605} (\bibinfo{year}{2004}).

\bibitem[{\citenamefont{{Howell} et~al.}(2005)\citenamefont{{Howell},
  {Tostevin}, and {Al-Khalili}}}]{How05}
\bibinfo{author}{\bibfnamefont{D.~J.} \bibnamefont{{Howell}}},
  \bibinfo{author}{\bibfnamefont{J.~A.} \bibnamefont{{Tostevin}}},
  \bibnamefont{and} \bibinfo{author}{\bibfnamefont{J.~S.}
  \bibnamefont{{Al-Khalili}}}, \bibinfo{journal}{J.\ Phys.\ (London) G}
  \textbf{\bibinfo{volume}{31}}, \bibinfo{pages}{S1881} (\bibinfo{year}{2005}).

\bibitem[{\citenamefont{Fortier et~al.}(1999)}]{For99}
\bibinfo{author}{\bibfnamefont{S.}~\bibnamefont{Fortier}} \bibnamefont{et~al.},
  \bibinfo{journal}{Phys. Lett.} \textbf{\bibinfo{volume}{461B}},
  \bibinfo{pages}{22} (\bibinfo{year}{1999}).

\bibitem[{\citenamefont{Winfield et~al.}(2001)}]{Win01}
\bibinfo{author}{\bibfnamefont{J.~S.} \bibnamefont{Winfield}}
  \bibnamefont{et~al.}, \bibinfo{journal}{Nucl. Phys A.}
  \textbf{\bibinfo{volume}{683}}, \bibinfo{pages}{48} (\bibinfo{year}{2001}).

\bibitem[{\citenamefont{{Cappuzzello} et~al.}(2001)\citenamefont{{Cappuzzello},
  {Cunsolo}, {Fortier}, {Foti}, {Khaled}, {Laurent}, {Lenske}, {Maison},
  {Melita}, {Nociforo} et~al.}}]{Cap01}
\bibinfo{author}{\bibfnamefont{F.}~\bibnamefont{{Cappuzzello}}},
  \bibinfo{author}{\bibfnamefont{A.}~\bibnamefont{{Cunsolo}}},
  \bibinfo{author}{\bibfnamefont{S.}~\bibnamefont{{Fortier}}},
  \bibinfo{author}{\bibfnamefont{A.}~\bibnamefont{{Foti}}},
  \bibinfo{author}{\bibfnamefont{M.}~\bibnamefont{{Khaled}}},
  \bibinfo{author}{\bibfnamefont{H.}~\bibnamefont{{Laurent}}},
  \bibinfo{author}{\bibfnamefont{H.}~\bibnamefont{{Lenske}}},
  \bibinfo{author}{\bibfnamefont{J.~M.} \bibnamefont{{Maison}}},
  \bibinfo{author}{\bibfnamefont{A.~L.} \bibnamefont{{Melita}}},
  \bibinfo{author}{\bibfnamefont{C.}~\bibnamefont{{Nociforo}}},
  \bibnamefont{et~al.}, \bibinfo{journal}{Phys. Lett.}
  \textbf{\bibinfo{volume}{516B}}, \bibinfo{pages}{21} (\bibinfo{year}{2001}).

\bibitem[{\citenamefont{{Crespo} et~al.}(2011)\citenamefont{{Crespo},
  {Deltuva}, and {Moro}}}]{Cre11}
\bibinfo{author}{\bibfnamefont{R.}~\bibnamefont{{Crespo}}},
  \bibinfo{author}{\bibfnamefont{A.}~\bibnamefont{{Deltuva}}},
  \bibnamefont{and} \bibinfo{author}{\bibfnamefont{A.~M.}
  \bibnamefont{{Moro}}}, \bibinfo{journal}{Phys. Rev. C}
  \textbf{\bibinfo{volume}{83}}, \bibinfo{pages}{044622}
  (\bibinfo{year}{2011}).

\bibitem[{\citenamefont{Lippmann and O'Malley}(1970)}]{Lip70}
\bibinfo{author}{\bibfnamefont{B.~A.} \bibnamefont{Lippmann}} \bibnamefont{and}
  \bibinfo{author}{\bibfnamefont{T.~F.} \bibnamefont{O'Malley}},
  \bibinfo{journal}{Phys. Rev. A} \textbf{\bibinfo{volume}{2}},
  \bibinfo{pages}{2115} (\bibinfo{year}{1970}).

\bibitem[{\citenamefont{Nunes}(1995)}]{Nun95}
\bibinfo{author}{\bibfnamefont{F.}~\bibnamefont{Nunes}}, Ph.D. thesis,
  \bibinfo{school}{University of Surrey} (\bibinfo{year}{1995}).

\bibitem[{\citenamefont{Nakamura et~al.}(1995)}]{Nak95}
\bibinfo{author}{\bibfnamefont{T.}~\bibnamefont{Nakamura}}
  \bibnamefont{et~al.}, \bibinfo{journal}{Nucl. Phys A}
  \textbf{\bibinfo{volume}{588}}, \bibinfo{pages}{c81} (\bibinfo{year}{1995}).

\bibitem[{\citenamefont{Palit et~al.}(2003)}]{Pal03}
\bibinfo{author}{\bibfnamefont{R.}~\bibnamefont{Palit}} \bibnamefont{et~al.},
  \bibinfo{journal}{Phys. Rev. C} \textbf{\bibinfo{volume}{68}},
  \bibinfo{pages}{034318} (\bibinfo{year}{2003}).

\bibitem[{\citenamefont{Fukuda et~al.}(2004)}]{Fuk04}
\bibinfo{author}{\bibfnamefont{N.}~\bibnamefont{Fukuda}} \bibnamefont{et~al.},
  \bibinfo{journal}{Phys. Rev. C} \textbf{\bibinfo{volume}{70}},
  \bibinfo{pages}{054606} (\bibinfo{year}{2004}).

\bibitem[{\citenamefont{Bertulani and Baur}(1988)}]{Ber88}
\bibinfo{author}{\bibfnamefont{C.~A.} \bibnamefont{Bertulani}}
  \bibnamefont{and} \bibinfo{author}{\bibfnamefont{G.}~\bibnamefont{Baur}},
  \bibinfo{journal}{Phys. Rep.} \textbf{\bibinfo{volume}{163}},
  \bibinfo{pages}{299} (\bibinfo{year}{1988}).

\end{thebibliography}

\end{document}